# Genotype networks in metabolic reaction spaces

*Areejit Samal[1,2,3], João F. Matias Rodrigues[4,5], Jürgen Jost[1,6], Olivier C. Martin[2,3,*,§], Andreas Wagner[4,5,6,7,*,§]*

[1] Max Planck Institute for Mathematics in the Sciences, Inselstr. 22, 04103 Leipzig, Germany
[2] Laboratoire de Genetique Vegetale du Moulon, UMR 0320/UMR 8120, Université Paris-Sud, 91190 Gif-sur-Yvette, France
[3] Laboratoire de Physique Théorique et Modèles Statistiques, CNRS UMR 8626, Université Paris-Sud, 91405 Orsay Cedex, France
[4] Department of Biochemistry, University of Zurich, Winterthurerstrasse 190, CH-8057 Zurich, Switzerland
[5] Swiss Institute of Bioinformatics, Quartier Sorge, Batiment Genopode, 1015 Lausanne, Switzerland
[6] The Santa Fe Institute, 1399 Hyde Park Road, Santa Fe, NM 87501, USA
[7] University of New Mexico, Department of Biology, 167 Castetter Hall, Albuquerque, MSC03 2020, USA
*These authors contributed equally to this work
§Correspondence author

Email addresses:
    AS: samal@mis.mpg.de
    JR: j.rodrigues@bioc.uzh.ch
    JJ: jost@mis.mpg.de
    OCM: olivier.martin@u-psud.fr
    AW: aw@bioc.unizh.ch

# Abstract

**Background**
A metabolic genotype comprises all chemical reactions an organism can catalyze via enzymes encoded in its genome. A genotype is viable in a given environment if it is capable of producing all biomass components the organism needs to survive and reproduce. Previous work has focused on the properties of individual genotypes while little is known about how genome-scale metabolic networks with a given function can vary in their reaction content.

**Results**
We here characterize spaces of such genotypes. Specifically, we study metabolic genotypes whose phenotype is viability in minimal chemical environments that differ in their sole carbon sources. We show that regardless of the number of reactions in a metabolic genotype, the genotypes of a given phenotype typically form vast, connected, and unstructured sets -- *genotype networks* -- that nearly span the whole of genotype space. The robustness of metabolic phenotypes to random reaction removal in such spaces has a narrow distribution with a high mean. Different carbon sources differ in the number of metabolic genotypes in their genotype network; this number decreases as a genotype is required to be viable on increasing numbers of carbon sources, but much less than if metabolic reactions were used independently across different chemical environments.

**Conclusions**
Our work shows that phenotype-preserving genotype networks have generic organizational properties and that these properties are insensitive to the number of reactions in metabolic genotypes.

# Introduction

The genotypes of biological systems form high dimensional spaces. A prominent example is that of proteins, where genotypes are strings of amino acids [1-2]. For a protein string of length $N$ and 20 proteinaceous amino acids, the genotype space consists of $20^N$ possible genotypes, an astronomically large number even for proteins of moderate length, The genotype spaces of RNA molecules and regulatory networks are similarly large [3-5]. If one imposes functional constraints on genotypes, the set of genotypes fulfilling these constraints is typically tiny compared to the entire space. In this work, we focus on a space of metabolic genotypes, and on the question how functional constraints structure this space.

Genotypes form phenotypes, observable traits of biological systems. Examples include the three-dimensional structure of proteins [6-7], secondary structure of RNA [5, 8-10] and the gene activity patterns of regulatory circuits [4, 11-14]. For some classes of biological systems, a mix of computational approaches and comparative data analysis has allowed systematic characterization of how individual genotypes map onto phenotypes [5, 15]. The genotype-to-phenotype maps arising in such works have several typical properties. First, any one phenotype is adopted by a vast number of genotypes. Second, these genotypes form large connected sets in genotype space. Specifically, it is possible to reach any genotype in such a connected set from any other genotype by a series of small genotypic changes, such as changes in individual amino acid sequences. Importantly, these changes leave the phenotype unchanged. Such a set of connected genotypes with the same phenotype is also referred to as a neutral network [5] or genotype network [16]. (We will be using the term "network" in two different senses here: on the one hand, we speak of a genotype network as a specific subset of genotype space. On the other hand, we refer to a metabolic network as a genotype belonging to a metabolic genotype space, as discussed below. We hope that this distinction will not confuse the reader.) Importantly, the term neutrality is not used here in the sense of fitness-neutrality as in the field of molecular evolution, but it just refers to invariance of a specific (and restricted) phenotype. Third, any one genotype network typically extends far through genotype space. Fourth, individual genotypes on any one genotype network typically have multiple neighbors with the same phenotype. Put differently, phenotypes are to some extent robust to small changes in genotypes (such as mutations). Finally, different neighborhoods of the same genotype network contain very different novel phenotypes.

Genome-scale metabolic networks are a class of biological systems that have received increasing attention in recent years [17-18]. They can be thought of as large assemblages of enzyme-catalyzed chemical reactions whose function is to produce all the small-molecule chemical compounds an organism needs to survive and reproduce in its environment. The major compounds include multiple amino acids, nucleotides, lipids, carbohydrates, and enzyme cofactors. Their relative proportions in an organism's biomass constitute the organism's biomass composition. The greater the rate at which an organism produces these compounds (in the right proportions), the faster the organism can grow and multiply. The metabolic network of any one organism can be thought of as existing in a vast genotype space of possible metabolic networks. Any one organism's genome encodes enzymes that catalyze some of these reactions. A simple representation of a metabolic network genotype $G$ uses a binary string of length $N$, e.g., $G = (b_1,...,b_N)$, where $N$ is the number of all enzyme-catalyzed chemical reactions that occur in the biosphere; this is illustrated in Figure 1a. Each position in the string representation of $G$ corresponds to one enzymatic reaction, and the necessary enzyme can be either present ($b_i=1$) or absent ($b_i=0$). In this framework, a metabolic genotype $G$ is a point in an $N$ dimensional hypercube (the genotype space) comprising $2^N$ metabolic genotypes.

For any one metabolic genotype, the computational approach of flux balance analysis (FBA) [17-19] can help determine whether the corresponding metabolic network can synthesize all major biomass components in a given chemical environment or medium. Flux balance analysis primarily uses information about the stoichiometry of enzymatic reactions in cellular metabolism to obtain a prediction for the steady state fluxes of all reactions and the maximum possible biomass synthesis rate of a metabolic network. The key underlying assumptions used in FBA are: (i) cellular metabolism operates at a steady state for a given environmental condition, wherein the concentrations of all internal metabolites and rates of all reactions are constant; (ii) the organism can adjust its metabolic fluxes -- rates at which individual reactions convert substrates into products -- to maximize its biomass growth flux. Flux balance analysis neglects regulatory properties of metabolic systems, and is thus concerned with the more fundamental constraints of biomass synthesis caused by the presence and absence of enzymes. In practice, regulatory constraints can often be overcome quickly in laboratory evolution experiments [20-21] and are thus temporary rather than fundamental obstacles to cell growth. The predictions of FBA and related approaches are often in good agreement with experimental results [22-23].

We here refer to the ability of a metabolic network genotype to synthesize a given biomass in a given environment as its metabolic *phenotype*. Except where noted otherwise, we will use the published biomass composition (i.e., its components *and* their proportions), of the bacterium *E. coli* [24], but our approach is not restricted to this organism. We call a genotype *viable* in a given chemical environment if it can synthesize biomass at some strictly positive rate in this environment.

Compared to other systems discussed above, little is known about the organization of metabolic phenotypes in metabolic genotype space [25-26]. In an earlier contribution [26] we demonstrated the existence of extended genotype networks for metabolic phenotypes that involved a large number of environments with many variable carbon sources. We also demonstrated that such genotype networks may facilitate evolutionary innovation in metabolism, that is, viability on novel carbon sources. A limitation of that earlier work is that we considered only metabolic networks within a narrow size range. Thus, we did not analyze the dependency of genotype space organization on network size systematically; nor did we estimate fractions of viable genotypes, and their dependency on network size. Here, we introduce an improved computational strategy to systematically address these issues. Our strategy, schematically illustrated in Figure 1b and explained below, is based on Markov Chain Monte Carlo sampling of the subspace of viable genotypes [27]. It allows us to study in detail the dependency of genotype network properties on the number $n$ of reactions catalyzed by a genotype. In particular, it reveals how the constraint of viability becomes more and more severe as $n$ decreases. The properties we examine include the typical and maximal distance of metabolic networks in a genotype network, their robustness to the removal of chemical reactions, and the dependency of these and other features on different chemical environments. With our approach and a suitable database of reactions, one can explore the space of genotypes having any desired phenotype.

# Results

**Viable genotypes get rarefied as the number of reactions *n* decreases**

We here explore the vast space $\Omega(n)$ of metabolic genotypes with a given number *n* of reactions. These reactions are taken from a "reaction universe" or "global reaction set" that comprises $N=2902$ reactions. Briefly, we obtained this reaction set by considering those reactions from the currently largest metabolic database KEGG [28], that fulfilled a number of criteria, such as being elementally balanced, and allowing nonzero biomass flux under steady-state conditions for at least one of our chemical environments ([26], see Methods for details). We are keenly aware that this set of reactions may comprise only a small fraction of the set of all chemical reactions in the biosphere. However, the metabolic genotype space $\Omega(N)$ it defines is vast and comprises many more reactions than a typical microbial metabolic network [17-18]. Our reaction set helps us to develop methods to explore this vast space, and to provide some intuition about its global properties.

We refer to a metabolic network genotype as any subset of reactions taken from this global set of *N* reactions. Such a genotype *G* can be represented as a binary string of length *N*, i.e., $G = (b_1,...,b_N)$, with each reaction *i* being either present ($b_i=1$) or absent ($b_i=0$). Note that even for modest *n*, $\Omega(n)$ is a huge space containing $N!/[n! (N - n)!]$ genotypes. A random genotype in $\Omega(n)$ corresponds to a random bit string subject to the constraint that there are exactly *n* bits set to 1.

For any one genotype, we use flux balance analysis (FBA) [17-19] to determine the maximum biomass flux obtainable for this genotype in a given chemical environment. Note that if a genotype produces a maximum biomass flux *F* under FBA, then adding further reactions can only increase *F*; as a consequence, in our framework, many properties will be monotonic in the number of reactions *n*. We here consider several minimal environments that differ in the sole carbon source they contain (see Methods). We call a genotype viable if its maximal biomass growth flux is strictly positive. We refer to the set or space of viable genotypes within $\Omega(n)$ as $V(n)$.

Perhaps the most basic question about $V(n)$ is how its size depends on *n*. A priori, it seems likely that metabolic networks with fewer reactions are less likely to be viable than networks of more reactions. Thus, one would surmise that $V(n)$ would decrease in size relative to the entire space $\Omega(n)$ as *n* decreases. To estimate this dependency, we need to estimate the probability $P_n=|V(n)|/|\Omega(n)|$ that a random genotype in $\Omega(n)$ is also in $V(n)$. This turns out to be difficult because $P_n$ can be very small. The problem can be circumvented through a decomposition approach. This approach takes advantage of the fact that $P_n$ can be written as a product of two factors, as we briefly explain now and detail further in Methods.

We call a reaction "essential" for a given viable genotype in a given chemical environment if its elimination ("knock-out") renders the genotype non-viable in that environment. We call a reaction "super-essential" in this environment if it is essential for the genotype that contains *all* reactions in the reaction universe. We note that a super-essential reaction will necessarily be essential for every viable genotype in this environment. The two factors in the calculation of $P_n$ are the probability that a genotype of a given number of reactions $n$ contains all super-essential reactions, and the probability that a genotype is viable given that it contains all super-essential reactions. The first factor can be determined analytically, as shown in Methods; we refer to it as an analytical prefactor. The second factor needs to be determined numerically. We estimated it through a sampling approach.

When applying this decomposition method to estimate the size of $V(n)$, we made the following observations (Figure 2). When $n$ is close to its maximum value, the viability constraint is very mild. Most genotypes are viable and $P_n$ is close to 1. As $n$ decreases, the probability of being viable decreases rapidly, and when $n$ gets too small, $P_n$ becomes too small to measure. In Fig.2 we show this behavior on a logarithmic scale, for three different minimal environments (glucose, acetate, and succinate). Figure S1 shows the same data on a linear scale (see Additional File 1). From these figures, it appears as if $P_n$ were independent of the sole carbon source (glucose, acetate, or succinate). However, there are differences among carbon sources, as we will discuss below; it is just that these differences are too small to be seen on the scale of Fig.2, for the range of $n$ that we were able to analyze.

Note that for genotypes that contain $n=2000$ reactions, that is, more than two thirds of the global reactions set, viable genotypes represent only a fraction of order $10^{-22}$ of all genotypes. In other words, the viability constraint is very severe even at this large number of reactions and it gets much more severe for smaller $n$. For completeness, Fig.2 also shows the analytical prefactor that is involved in our estimates of $P_n$; note that it captures most of the trend of the data. This means that the constraint of including super-essential reactions is far more stringent than the constraint of being viable given that one includes such reactions. In other words, most of the likelihood of being viable comes from the likelihood of containing all super-essential reactions. Furthermore, this likelihood scales approximately exponentially in $n$: each removal of a reaction decreases the probability of containing all super-essential reactions by approximately a constant factor. Finally, the fact that the fraction $P_n$ of viable genotypes is a convex curve (Fig.2) indicates that the effect of successive reaction removals on viability becomes more and more severe.

**MCMC sampling of viable genotypes**

We next wanted to study the properties of *random* genotypes in $V(n)$, *i.e.*, metabolic networks sampled uniformly from the genotype space $\Omega(n)$, and subject only to the constraint that they are viable. For instance, one can ask what is the typical mutational robustness of such genotypes, or how frequently do they contain specific reactions of the global reaction set. The answers to such questions can provide valuable null-hypotheses and points of contrast with organismal metabolic networks like that of *E. coli*.

Uniform sampling of viable genotypes is difficult, given that $V(n)$ can be so much smaller than $\Omega(n)$. Nevertheless, this task can be tackled as follows. For $n$ greater than 2500, one can easily generate random genotypes and include them in a statistical analysis if and only if they are viable. For smaller values of $n$, that kind of sampling becomes very inefficient and it is better to use the Markov Chain Monte Carlo (MCMC) method, a widely used approach for sampling large and high dimensional spaces [27]. The MCMC method produces a sequence of genotypes forming a chain, the term "chain" coming from the property that the (k+1)$^{th}$ element of the sequence is generated from the k$^{th}$ one using a probabilistic transition rule. At each transition, one proposes a small modification to the current genotype; if this modified genotype is viable, one accepts it as the next genotype of the sequence, otherwise the next genotype is identical to the current genotype. In our work, the modification introduced at each transition step is a reaction swap. That is, each modification adds one reaction and removes another reaction, so as to keep *n* constant. The MCMC thereby produces a walk in the subspace of viable genotypes of *n* reactions as illustrated in Figure 1b; an overview of the different steps of the algorithm is displayed in the flow chart of Figure S2 (*cf.* Additional File 1) and technical details are given in the Methods. Of particular relevance is that, in the limit of long walks, this approach can be shown to sample *uniformly* the space of viable genotypes that are accessible when starting with the first genotype of the MCMC. Whether or not this accessible space is the whole space *V(n)* is difficult to determine. We thus take a practical approach, and simply focus on properties of genotypes in the accessible space.

To start the Markov chain, a first viable genotype is necessary. We generated such a genotype as follows. For the *E. coli* metabolic network genotype, we determined those reactions that have nonzero flux in a steady-state flux distribution that yields a nonzero biomass growth flux in a given chemical environment. For the environments we studied here, this number of reactions with nonzero fluxes is smaller than 300. We then start with this set of reactions and add randomly other reactions until we have reached a metabolic network with exactly *n* reactions. The corresponding genotype is viable in the given environment, it has the right number of reactions, and we can thus use it as the first element of the MCMC sequence.

With this algorithm, we can sample the accessible space uniformly; any observable such as the mean mutational robustness can be estimated from a sample; and the estimate's uncertainty (error bar) can be computed as explained in Methods. We find that the procedure is relatively efficient, as judged by either the chain's autocorrelation time τ or by the acceptance rate of each genotype modification (*cf.* Figs. S3 and S4 in Additional File 1, see also Methods). From Figure S3, one can see that τ≈2000, so the MCMC generates uncorrelated genotypes every few thousand swaps (see Additional File 1).

**Mutational robustness increases as *n* increases and has a narrow distribution**

Given a chemical environment and a viable metabolic network genotype *G*, we define the mutational robustness $R_\mu$ of *G* as the fraction of its reactions that are not essential in that chemical environment. Thus, $R_\mu$ is simply the probability that a metabolic network remains in the viable space under one random reaction deletion. Qualitatively, one expects that for large *n*, most genotypes in *V(n)* will have a high mutational robustness Nevertheless, there are close to 100 super-essential reactions for each of the chemical environments we study. Removal of any of these reactions will yield a non-viable genotype. This means that even when *n* is at its maximum value, $R_\mu$ must be below 1. Furthermore, one expects $R_\mu$ to decrease with *n*, perhaps quite steeply. To study this dependency on the number of reactions in a quantitative way, we used the MCMC approach for multiple values of *n* to generate samples of 1000 random viable genotypes and determined their values of $R_\mu$. Figure 3 shows the *mean* mutational robustness of genotypes in *V(n)* as a function of *n* for three different minimal environments using the *E. coli* biomass composition. We see that $R_\mu$ begins to decrease steeply near *n*=1000. By the time *n*=300 is reached, the mean mutational robustness has reached a value close to 0.15. This is a low but not tiny value of robustness. (We note parenthetically that $R_\mu$ is of the same order of magnitude as the swap acceptance rate of the MCMC. Values of $R_\mu$ that are not vanishingly small amount to MCMC sampling that is not very inefficient.)

We also studied the *distribution* of $R_\mu$ for random viable genotypes at multiple values of *n*. For a first series of estimates, we used the *E. coli* biomass composition [24]. An illustrative case is shown in Figure 4a for the glucose environment when *n*=831, the value of the number of reactions from our database that are contained in the *E. coli* metabolic network. The cases for acetate and succinate environments are shown in Figs. 4b and c. The horizontal axes of these figures show that the width of the distribution spans only 2 percent of the possible range between zero and one. In other words, $R_\mu$ varies very little among random viable genotypes. It is instructive to compare this low variance in robustness to that predicted by a simple probabilistic argument, where each reaction contributes independently to mutational robustness. Suppose that each reaction in each genotype in *V(n)* has the same probability *r* of being non-essential. This probability *r* would then be given by the mean of $R_\mu$, which is around 0.715 for the genotypes of Figure 4a. If all reactions contribute independently to mutational robustness, then the robustness of any one genotype is the average of *n* binary numbers that adopt a value of 1 with probability *r*. For the large values of *n* we consider, this quantity effectively follows a normal distribution with mean *r* and variance *r(1-r)/ n*. However, to make this argument more powerful, it is best to note that super-essential reactions do not contribute to the variance of $R_\mu$. Thus, we only need to consider the non-essential reactions. If *s* is the number of super-essential reactions, the expected variance of mutational robustness then becomes $p(1-p)(n-s)/n^2$ where *p=nr/(n-s)*. Figs. 4a-c show the distribution of mutational robustness predicted by this argument as dashed lines. For example, in the case of the genotypes of Figure 4a, this framework yields a value of the variance of $1.6 \times 10^{-4}$, which is close to the variance of $1.1 \times 10^{-4}$ we estimate through sampling. In sum, these observations suggest that the distribution of $R_\mu$ is narrow simply because the number of terms that contribute to $R_\mu$ is large. Figure S5 corroborates this argument, displaying the distribution of $R_\mu$ at *n*=700 along with the normal distribution predicted by the probabilistic argument (see Additional File 1). Again, this argument predicts the actual distribution well.

We next turn to the mutational robustness of *E. coli* itself and ask whether it is unusually high or low. Specifically, we introduce the null hypothesis $H_0$ that *E. coli's* robustness is indistinguishable from that of a random genotype in *V(n)*. The mutational robustness $R_\mu$ of *E. coli* has values of 0.745, 0.729 and 0.735 in glucose, acetate, and succinate. For all three environments, these values are in the upper tail of the distribution of $R_\mu$ for randomly sampled viable genotypes in *V(n)*. Based on the location of *E. coli's* robustness in these distributions, we can reject $H_0$ at *p*-values of 0.001, 0.04, and 0.02 for the glucose, acetate, and succinate environments, respectively. *E. coli* thus appears to be atypically robust. We contrast this observation with that of Figure 2a in Ref. [26] where an analogous MCMC study led to a different distribution for $R_\mu$ under $H_0$, and to the conclusion that *E. coli* was not an outlier. The source of the different conclusions between these two studies is this:. In our MCMC approach, the number *n* of reactions is exactly fixed, whereas it was allowed to vary modestly over a range of values in Ref. [26]. That range contained mainly values larger than 831, and thus generated a sample of random viable genotypes with higher robustness. Clearly, it is better to use the present MCMC method where *n* can be set to its value in *E. coli*.

One might argue that it is inappropriate to call a reaction non-essential if only a tiny biomass growth flux is realizable when the reaction is absent. Put differently, our results might change if we required that removal of a reaction did not reduce biomass growth flux below some threshold. To assess whether this is the case, we used the MCMC method described above to sample only those genotypes in *V(n)* that have a large biomass flux. Specifically, we sampled the space of viable genotypes for which the biomass flux is at least as large as the *in silico E. coli* biomass flux. The resulting distribution of $R_\mu$, along with the *E. coli* mutational robustness is displayed in Fig.4d. The figure shows that the distribution of mutational robustness is essentially the same whether one imposes this more stringent requirement on flux, or the much looser constraint of any strictly positive biomass flux. *E. coli* thus remains an outlier with significantly high robustness also in the more stringent comparison.

These results were all obtained using the published proportions at which *E. coli* biomass constituents [24] occur in the *E. coli* cell. We next asked whether the features we found (narrowness of the distribution of $R_\mu$, *E. coli* being an outlier) are sensitive to this choice. To find out, we repeated the above analysis, but changed the proportion of each biomass constituent randomly by up to 20%. All the above conclusions hold for this new formula as well; this is illustrated in Figs. S6 a-d (see Additional File 1).

**Mean and maximum distances between genotypes in *V(n)***

We showed above that viable genotypes form a tiny fraction of the space of all possible genotypes, even for numbers of reactions that are modestly large. For instance, at *n*=2000, the viable genotypes form a fraction of about $10^{-22}$ of all genotypes. This fraction would decrease further as we approach reaction numbers close to that of *E. coli*. Is this viable space *V(n)* concentrated in a small region of the total space *Ω(n)*? In other words, are most viable genotypes very similar to one another? To address this question, we used the Hamming distance $D_H$ of the bit strings associated with two metabolic genotypes as a measure of genotype distance. In *Ω(n)*, this Hamming distance can range from a value of zero to *2 min(n,N-n)*.

We first estimated the distribution of $D_H$ between genotypes in *V(n)* from our samples of 1000 random viable genotypes. The average of this distribution is shown in Fig.5a for different values of *n*, and for the glucose, acetate and succinate environments. To have a comparison benchmark, we also show the mean of $D_H$ for bitstrings (genotypes) chosen at random from *Ω(n)*. This quantity is easily calculated analytically, and obviously does not depend on the environment. Its value is *2 n (N-n)/N*. Figure 5a shows that when *n* is close to *N*, this mean value of $D_H$ is close to the maximum distance in *Ω(n)*.

Figure 5a shows that the mean Hamming distance in *V(n)* is nearly independent of the chemical environment. Furthermore, for *n* greater than *N/2*, random viable genotypes are not much more similar than random genotypes in *Ω(n)*. Thus in this regime the viability constraint is nearly invisible. In contrast, when *n* is smaller than *N/2*, the constraint of viability leads to genotypes that are significantly more similar than random genotypes. For instance at *n*=500, the mean Hamming distance is about 42% higher in *Ω(n)* than in *V(n)*.

After having estimated the typical distance of random viable genotypes, we turned to the *maximum* possible distances of such genotypes. This distance can be thought of as the "span" or diameter of *V(n)*. To estimate this quantity, we generated two parallel random walks in *V(n)* and at each step tried to maximize the Hamming distance between them. Specifically, we followed simultaneously two genotypes in a modified MCMC approach, where the criterion for accepting a genotype modification involved not only viability, but also the requirement that the Hamming distance to the other genotype be non-decreasing. During a typical execution of such a random walk, the distance between the two genotypes rises and then saturates. We carried out random walks of up to $10^6$ steps, and terminated any such walk if the distance between the two genotypes had not increased in the last third of the walk. We repeated this procedure 5 times for each value of *n* and each environment, and kept the largest distance identified in these 5 replicates as a final estimate. We note that this procedure provides a lower bound of the actual maximum distance between genotypes in *V(n)*. The results are shown in Fig.5b for the three environments, along with the maximum distance in *Ω(n)* namely *2 min(n,N-n)*. Qualitatively, we obtained a pattern similar to that of the average distance (*cf.* Figure 5a). For example, the viability constraint is invisible for large *n*, but it strongly affects the maximum genotype distance for *n < N/2*; not surprisingly, this viability constraint becomes more marked as *n* decreases to lower values. However, the magnitude of this effect is smaller than for the average distance discussed above. For example, at *n*=500, the maximum distance is 29% higher in *Ω(n)* than in *V(n)*, while for the mean distance the corresponding percentage is 42%.

**Clustering analysis of genotypes in *V(n)***

The space *Ω(n)* is a subset of a 2902-dimensional hypercube whose members have exactly *n* bits set to one. Thus, it is a hyperplane section of the hypercube. The distribution of distances in this space is unstructured and homogeneous. This may not be the case for the viable subspace *V(n)*. For example, some of the genotypes in *V(n)* may present different "types" of solutions to the problem of producing a viable genotype. These different types of solutions could manifest themselves in different clusters of genotypes within *V(n)*. To determine whether this is the case, we have performed a principal component analysis, using the bitstring representations of 1000 random viable genotypes. The results for the first two principal components are given in Fig.6 for the glucose chemical environment, and for a number *n*=831 of reactions equivalent to that in the *E. coli* reaction set. The first two principal components explain less than 1 percent of the variance in the genotype differences. Similar observations hold for succinate and acetate environments (results not shown). For completeness, the figure also shows the localization of the *E. coli* genotype. In this analysis, its location is not atypical, in contrast to our observations for mutational robustness. We supplemented this analysis by a hierarchical clustering analysis of the same 1000 genotypes, which also did not reveal any grouping of genotypes (*cf.* Figure S7 in Additional File 1). In sum, these analyses suggest that viable genotypes of this size form a "cloud" in genotype space with little internal structure and heterogeneity.

**Use of reactions by genotypes in *V(n)***

Previous work [26] showed that not all reactions are equally likely to occur in random genotypes that are viable in a given chemical environment. In our next analysis, we analyzed the nature of these differences among reactions more systematically. Clearly, at one extreme, super-essential reactions are present in every viable genotype. At the other extreme, there may be many reactions that occur with a low but non-negligible probability in random viable genotypes. For example, a random viable genotype may contain superfluous reactions. (Note that the addition of such reactions in and by itself cannot cause loss of viability in our framework: if a genotype produces a maximum biomass flux $F$ under FBA, then including extra reactions cannot decrease $F$.) In addition, there may be intermediate cases where reactions are present often but without being super-essential. Fig.7 shows a rank histogram indicating the occurrence of all the reactions in the global reaction set for random viable genotypes in a glucose minimal environment. To compute this histogram, we first determined how often each of the 2902 reactions occurred in a sample of 1000 genotypes from *V(831),* that is, in random viable genotypes with as many reactions as the *E. coli* genotype in our framework. For any one reaction, the "occurrence" plotted on the vertical axis of the figure is the fraction of genotypes in this sample that contain the reaction. Then we ordered the reactions according to their occurrence, assigning rank "1" for the reaction with the highest occurrence and rank 2902 for the reaction with the lowest occurrence. All super-essential reactions have an occurrence of 1.0 and contribute to the horizontal plateau on the left of the rank histogram, similar to previous observations [26]. We also see a much broader plateau on the right with occurrence values of approximately 0.2. The two plateaus are connected by a graded slope, corresponding to reactions whose occurrence decreases continuously from 1.0 to 0.2. To study this intermediate region of the rank histogram, we first fitted this histogram to two classes of functions $f(r)$: a function that represents occurrence as a constant plus an exponential function of the rank ($f(r) = a + b\ exp[-c\ r]$), and a function representing occurrence as constant plus a power function of the rank ($f(r) = a + b / r^c$ ). The power function provides a visibly better fit in Figure 7. This difference can be analyzed quantitatively by calculating the coefficient of determination $R^2$ which represents the variance explained by each functional relationship. We find that $R^2=0.96$ for the exponential function and $R^2=0.99$ for the power law. Note that power laws like this arise in other situations with rank histograms [29].

     We next asked whether reactions that are neither super-essential nor on the right hand side plateau of Figure 7 are often essential reactions. To do so, we returned to our sample of 1000 viable genotypes, and computed for each genotype of this sample the maximum flux through each reaction using the flux variability approach [30]. If this flux was equal to zero for a reaction, we considered the reaction to be "blocked" [31-32]. We also determined for each genotype its essential reactions. We thus obtained for each reaction the frequency at which it is essential and the frequency at which it is blocked (in our sample of 1000 genotypes). Figures 8 (a) and (b) are scatter plots of these quantities as a function of the occurrence of each reaction. Clearly, reactions of low occurrence tend to be blocked, whereas reactions with high occurrence are often essential. In summary, the smooth transition region in the rank histogram is a transition from essential or near essential reactions to reactions that cannot possibly have a function in a given genotype and environment.

The scalar value of reaction occurrence is essentially a one-dimensional projection of a high dimensional pattern of the distribution of reactions among random viable networks. This high-dimensional pattern may be structured in other, non-obvious ways. To find out whether this is the case, we carried out a principal component analysis. To this end, we organized the 1000 sampled genotypes into a matrix, such that each row corresponds to the bit string associated with one genotype. Then, each column corresponds to some reaction $R$, and comprises a bit string of length 1000 that contains a '1' in each row where $R$ is part of the corresponding genotype. Figure S8 shows the results of the principal component analysis carried out on these bit-strings reflecting reaction occurrence (see Additional File 1). This analysis used the same 1000 random viable genotypes in $V(831)$ that we analyzed above. The figure shows the first and the second principal component, which explain ~14% and ~1% of the variance in the data, respectively. The first principal component corresponds well with the rank of the reaction as defined by its occurrence. To visualize the association between the first axis and reaction rank, we have colored the reactions according to their rank (red for ranks close to 1, indigo for ranks close to 2902). The data is clearly heterogeneous, resembling a "comet" with a dense head on the left and a spread-out tail on the right. The comet's head is formed mostly by blocked reactions, while the tail of the comet is enriched with essential reactions. In contrast, the second principal component displays no particular clustering. In sum, reactions that occur in different random viable genotypes show no obvious statistical patterns beyond their rank. This rank strongly correlates with reaction essentiality and blockage.

**Random genotypes show highly correlated viability in different environments**

Although each chemical environment $i$ has a specific viable space $V_i(n)$, some genotypes in $V_i(n)$ must be viable for multiple environments, in particular at large $n$. To what extent do the viable spaces for different environments overlap? What are the properties of genotypes in their intersection? To ask these and other questions, we extended our computational approach to multiple environments.

To begin with, we wanted to know whether being viable in one environment improves the chances of being viable in another environment. We therefore estimated the conditional probability $P(2|1)$ of being viable in a given environment 2 under the condition of being viable in another environment 1. As we saw earlier, the probability of being viable in one environment is tiny for genotypes with fewer than $n=2500$ reactions. If the probabilities of being viable in two different environments were independent from one another, $P(2|1)$ would be equal to $P(2)$. We estimated $P(2|1)$ by first uniformly sampling the set of viable genotypes in environment 1 using the MCMC method. From the 1000 genotypes thus generated, we determined the fraction of genotypes that are also viable in environment 2. In the case of independence, this fraction should be close to the value of P(2), because then $P(2|1)=P(2)$ In contrast, we find that $P(2|1)$ is orders of magnitude higher. This is illustrated in Figure 9 for all six possible pairwise combinations of three environments (glucose, succinate and acetate). It is useful to compare the data shown there to that of Figure 1. For instance at $n=2100$, if $P(2|1)$ were equal to $P(2)$, Figure 9 should show a probability $P(2|1)$ of about $10^{-18}$, but the actual value exceeds 0.9, regardless of the environment pair considered. Similarly, for $n=1000$, Figure 9 shows that $P(2|1)=10^{-2}$ or greater, whereas $P(2)$ is below our detection limits for genotypes with this number of reactions.

The large association between viability in different environments we observed should arise if genotypes viable in different environments require nearly identical super-essential reactions. The reason is that the probability that a genotype contains all super-essential reactions is the dominant determinant of the probability that the genotype is viable (Figure 2). In the extreme case where the super-essential reactions for two environments are *identical*, only the remaining, minor determinants of the probability to be viable come into play. In consequence, the conditional probability $P(2|1)$ can be large. For the environment pairs we study here, the set of super-essential reactions are in fact nearly identical: all three environments share the same 99 super-essential reactions, and only acetate has one additional super-essential reaction. Thus the strong association between the viability in different environments is no surprise. We emphasize that this does not mean that the sets $V(n)$ of genotypes viable in different environments are nearly identical. If that were the case, the conditional probabilities displayed in Fig. 9 would be close to one. Instead, these probabilities become very small at our lowest values of $n$.

In sum, the sets of viable genotypes $V_i(n)$ for different environments are far more similar than expected by chance. Nonetheless, their intersections contain only a small fraction of any one set $V_i(n)$ as $n$ decreases.

**Different environments constrain viability to different degrees**

The MCMC method allowed us to estimate the conditional probability $P(2|1)$ that a genotype is viable in environment 2 given that it is viable in environment 1. We now show that it also allows us to estimate the relative sizes of the viable spaces $V(n)$ for different chemical environments. We define the following probability

$$P(i) = |V_i(n)| / |\Omega(n)|$$

for chemical environment $i$, where $V_i(n)$ is the set of genotypes with $n$ reactions that are viable in environment $i$. In addition, we define $P(i,j)$ to be the fraction of genotypes with $n$ reactions that belong to both $V_i(n)$ and $V_j(n)$. Then by MCMC sampling of $V_i(n)$ we can estimate the fraction $P(i,j)/P(i)$ as outlined in the previous subsection. Analogously, by sampling $V_j(n)$, we can also estimate the fraction $P(i,j)/P(j)$. The ratio of these estimates then yields the ratio $P(j)/P(i)$. In other words, this approach can tell us whether fewer genotypes are viable in one chemical environment than in another. This ratio is shown in Figure S9 for environments defined by the three carbon sources we considered here (see Additional File 1). Glucose has the largest set of viable genotypes, followed by succinate and acetate. As $n$ decreases, the differences in the sizes of the $V_i(n)$ for these environments become more pronounced, because constraints on viability become more severe in that limit.

The concept of a probability $P(i,j)$ of being viable in both environments $i$ and $j$ can be extended to any number of environments by defining appropriate ratios $P(i,j,...k)$. Although the MCMC approach does not allow us to estimate these probabilities directly, it does provide estimates for *ratios* of the quantities $P(i)$, $P(i,j)$, $P(i,j,k)$, *etc*., as just discussed. And just like we found that glucose has the largest space $V(n)$ of viable genotypes of the three carbon sources we consider, we can ask which *pair* of carbon sources has the most viable genotypes associated with it. Fig. S10 shows that this pair is the succinate-acetate pair, while the pairs that contain glucose are about equivalent (see Additional File 1). Because the MCMC approach only provides ratios, the figure shows the three *P(i,j)* divided by a common normalization factor, namely *P(*Glucose*)*. This procedure allows a direct comparison of the relative values of the *P(i,j)*.

For completeness, we show in this same figure the quantity *P(i,j,k)/P(i)*, that is, the probability *P(i,j,k)* of being viable on all three carbon sources divided by the probability *P(i)* of being viable on environment *i* (glucose in this figure). Such data can tell us to what extent increasingly complex nutritional requirements affect viability.

To continue in this direction, we can ask whether being viable in one environment leads to an appreciable probability of being viable in *many* additional environments. To test this, we considered 88 aerobic minimal environments that differ in their carbon sources, as listed in Table 1 of [33], and asked in how many of these environments a random viable genotype in $V_i(n)$ will be viable. In Figs. S11a-c we show the histogram of the number of these additional environments, with *n* set to its *E. coli* value (see Additional File 1). Genotypes viable in all three "reference" (glucose, succinate and acetate) environments exhibit a bell shaped distribution of the number of environments in which they are viable. This indicates that typical genotypes in $V_i(n)$ will be viable in more than just their reference environment.

**Robustness in different environments is highly correlated**

We next extended our analysis of robustness to multiple environments. We sampled 1000 random genotypes that belonged to *both* $V_i(n)$ and $V_j(n)$, and measured for each its mutational robustness, defined as the fraction of reactions that were non-essential in *both* environments *i* and *j*. In Fig.S12a we show as a function of *n* the mean robustness found for the three choices of environment pairs involving glucose, succcinate or acetate as sole carbon sources (see Additional File 1). Just as for the single environment case, we see that the mean robustness increases as *n* increases and that the lowest values are not extremely small. We extended this analysis to the case of three environments as shown in Fig.S12b which shows the same trends (see Additional File 1).

The use of 1000 genotypes viable on multiple environments allows us to estimate also the *distribution* of mutational robustness in multiple environments. Not surprisingly, just as for a single environment, the distribution is narrow. We illustrate this in Figure S13a and b for networks of *n*=831 reactions, as in the *E. coli* metabolic network (see Additional File 1). Figure S13a shows data for the pair glucose-succinate, and Figure S13b for the triplet glucose-succinate-acetate (see Additional File 1). Both panels alsoindicate the mutational robustness of *E. coli*; it is a clear outlier, in fact even more so than in a single environment (*cf.* Figure 4).

Finally, we asked whether there are *trade-offs* in genotype robustness for these multiple environments. For example, a genotype highly robust to reaction removal in one environment may have low robustness in a second environment. However, this is not the case. Figures 10a-c are scatter plots based on 1000 random genotypes viable in pairs of environments. Here, robustness in a given environment is again defined as the fraction of non-essential reactions in *that* environment. For all three environment pairs, Figures 10a-c show a strong positive association between robustness in two different environments rather than a trade-off. We note that this association cannot be explained by the fact that the super-essential reactions are shared between environments. The reason is that these reactions cannot contribute any *variability* to mutational robustness.

**Clustering of viable genotypes in multiple environments**

In a final analysis, we asked whether genotypes in $V_i(n)$ can be distinguished from genotypes in $V_j(n)$, merely by their different "locations" in the genotype space $\Omega(n)$. To answer this question, we performed a principal component analysis of three pooled sets of data: 1000 random genotypes in $V_i(n)$, another 1000 random genotypes in $V_j(n)$, and yet another 1000 random genotypes in the intersection of $V_i(n)$ and $V_j(n)$. In Figure S14, we show the first and second principal components resulting from this global analysis, where *i* corresponded to glucose and *j* corresponded to succinate (see Additional File 1). The figure also shows the center of mass for each of the three samples. We observe three clouds of genotypes, one for each set. These clouds overlap, but they are also different in a statistically significant way. To demonstrate this claim, we performed a permutation test as follows. First, we defined $D^*$ as the sum of the three distances between the three centers of mass of the samples. Second, we generated $10^4$ random permutations of the points in the clouds, thereby shuffling their assignment to the three clouds. For each shuffle *k*, we determined the new centers of mass and computed $D_k$ as the sum of their three distances. The p-value of the hypothesis that there is no clustering of genotypes according to their viability set is then estimated as the fraction of the shuffles having $D_k$ greater than $D^*$. In our test, we found no shuffle satisfying that inequality, so our estimate of the p-value is less than $10^{-4}$. The same analysis was performed for the other two pairs of environments, but the pair shown has the clearest separation of the clusters.

# Discussion and Conclusion

In sum, our analysis has revealed several characteristic features of the genotype space of metabolic networks. First, the probability that a random metabolic network in genotype space is viable (produces a strictly positive biomass growth flux) decreases dramatically as its number of reactions *n* decreases from the number *N* of reactions in a reaction "universe". Most of this decrease is caused by the fact that smaller metabolic networks are less likely to contain all super-essential reactions, reactions that any network must contain to be viable in a given environment. Second, the robustness of random viable metabolic networks to the removal of random reactions shows a very narrow distribution. This observation is a simple consequence of a law of large numbers, because many individual reactions contribute to the viability of a reaction network. In all environments we studied, the *E. coli* metabolic network is significantly more robust than random viable networks. Third, we showed that random viable networks typically contain very different sets of reactions. Specifically, their genotype distance is not much below that of metabolic networks chosen at random from genotype space, regardless of their viability. Viable genotypes having few reactions are exceptions to this rule, because they are appreciably more similar to each other than random genotypes.

     Fourth, the maximum genotype distance between viable metabolic networks is almost as large as the diameter of genotype space itself, that is, to the maximum distance among all genotypes, regardless of their viability. Because of the method we used to identify the typical and maximum genotype distance, we know that genotypes at these

distances can be connected through individual reaction changes that do not affect viability. In other words, viable genotypes form genotype networks that extend far through genotype space.

Fifth, there is remarkably little structure within a given genotype network. A principal component analysis and hierarchical clustering of our samples of random viable genotypes detects no clustering of genotypes that might indicate different types of viable metabolic networks. Importantly, none of these observations depends strongly on the specific carbon source we used in our minimal chemical environments. However, there are differences between the number of genotypes viable on different carbon sources. For example, the space of viable genotypes is smaller for acetate than for glucose. Also, fewer genotypes are viable simultaneously on two or three of our carbon sources than on just one carbon source. Our Markov Chain Monte Carlo sampling approach allows us to estimate the relative sizes of these viable spaces.

We next discuss two potential caveats to our analysis. First, in order to preserve reaction numbers, we used the decidedly non-biological choice of swapping reactions to carry out random walks in our MCMC exploration of genotype space. In contrast, during the evolution of genome-scale metabolic networks, individual reactions are eliminated from metabolic networks through loss-of-function mutations in their enzyme-coding genes; reactions are added through horizontal gene transfer [34-35] or, more rarely, via the evolution of a protein with new catalytic function within an organism. In this regard, we note that every reaction swap can be viewed as an addition of a reaction (which always maintains viability) followed by a reaction deletion that preserves viability. In other words, a reaction swap can be decomposed into a succession of biologically meaningful genotypic changes; the connectivity of genotype networks found in our swap approach thus holds also for single additions and deletions of reactions.

Our knowledge of the biochemical reaction universe will undoubtedly grow in the future, which raises the question of how sensitive our observations are to the addition of reactions to this universe. In this regard, it is worth highlighting that the added reactions are unlikely to affect viability in the common carbon sources we studied. For such carbon sources, the super-essential reactions are known, and our current knowledge of the reaction universe already allows for an astronomical number of viable genotypes. We suspect that key quantities, such as the large diameter of genotype networks or their typically high robustness to mutations will thus increase rather than decrease as more and more "accessory" reactions become known. These arguments do not necessarily apply to more complex or exotic chemical environments. We leave the exploration of such environments to future work.

The work we reported here suggests that viability in the *E. coli* metabolic network is significantly more robust to random reaction removal than in random viable genotypes for the three environments we studied. This high robustness persists if we require that a viable genotype does not just have positive biomass growth flux, but also a high biomass growth flux, and also if we change the stoichiometry of the biomass composition randomly by up to 20 percent. We note that because of the very narrow distribution of robustness among random viable genotypes, this significant difference translates into a modest decrease in the absolute number of essential reactions in *E. coli*. For example, in a glucose minimal environment, random viable genotypes in our sample have a mean number of 237 essential reactions, whereas *E. coli* has 212 essential reactions, *i.e.,* 25 fewer than the mean. We cannot exclude that future improvements of the *E. coli* network annotation might reduce this significant difference to random viable metabolic networks, and thus our observation of significantly high *E. coli* robustness should be interpreted with caution. Earlier work by some of us [26] had suggested that the robustness of the *E. coli* network was not significantly different from random viable networks. However, in that work we had allowed the number of network reactions in an MCMC exploration of metabolic network space to fluctuate by a modest amount. Because of the modest number of differences in essential reactions between *E. coli* and random viable metabolic networks, these fluctuations were sufficient to obscure significant differences in robustness. Our improved MCMC approach leaves reaction numbers strictly constant and allows us to reveal these differences.

In conclusion, our work shows that phenotype-preserving genotype networks that extend far through metabolic genotype space are not just peculiarities of metabolic networks with a given size. They are generic organizational properties of metabolic genotype space, and share important features of genotype networks in other classes of systems, such as proteins, RNA, and regulatory circuitry [5, 9, 36].

## Materials and Methods

### Flux Balance Analysis (FBA)

Flux balance analysis (FBA) is a constraint-based modeling approach that primarily uses the information about the stoichiometry of all enzymatic reactions in cellular metabolism to obtain a prediction for the steady state fluxes of all reactions and maximum biomass yield of the organism [18-19]. The information about the stoichiometry of metabolic reactions is encapsulated in the stoichiometric matrix **S** of dimensions $m \times n$, where $m$ denotes the number of metabolites and $n$ denotes the number of reactions. In any metabolic steady state, different metabolites achieve a mass balance where the vector **v** of reaction fluxes satisfies the equation

$$\mathbf{Sv} = 0 \quad (1)$$

representing the stoichiometric constraints and the requirement of mass conservation in the steady state. For genome-scale metabolic models, the above equation typically leads to an under-determined system of linear equations, and a large solution space of allowable fluxes. The size of this space can be reduced by incorporating thermodynamic constraints associated with irreversible reactions, as well as flux capacity constraints, which limit the maximum flux through some or all reactions. Linear programming (LP) can then be used to find a set of flux values -- a point in the space of allowable solutions -- that maximize a certain biologically relevant linear objective function $Z$, which is usually chosen to be the biomass growth flux. The LP formulation of the FBA problem can be written as:

$$\max Z = \max\{\mathbf{c}^T\mathbf{v} \,|\, \mathbf{Sv} = 0, \mathbf{a} \leq \mathbf{v} \leq \mathbf{b}\} \quad (2)$$

where the vector **c** corresponds to the objective function, and vectors **a** and **b** contain the lower and upper limits of different metabolic fluxes contained in **v**.

**Global reaction set**

In this work, we have used a set of 5870 reactions compiled earlier by two of us (J.R. and A.W.) through merging of the Kyoto Encyclopedia of Genes and Genomes (KEGG) LIGAND reaction database [28] with the *E. coli* metabolic model iJR904 [24], followed by appropriate pruning to exclude generalized polymerization reactions [26]. Of the 5870 reactions in this hybrid database, 3369 are irreversible and 2501 are reversible reactions. We took the set of 143 external metabolites contained in the *E. coli* iJR904 model to be the set of possible uptake and secreted metabolites in the metabolic network. We have used an objective function $Z$ (Eq. 2) that requires synthesis of those *E. coli* biomass compounds defined in the iJR904 model [24]. In this function, we also used the proportions of these compounds defined in [24].

Large-scale metabolic networks typically have certain dead-end reactions; these can only have zero flux for every investigated chemical environment, under any steady state condition with nonzero biomass growth flux. Such reactions have been referred to as "blocked" in the literature and cannot contribute towards the steady state flux distribution [31-32]. We found that 2968 of the 5870 reactions would be blocked under all environmental conditions that we examine in this study. We excluded this set of 2968 blocked reactions from the hybrid database of 5870 reactions, which led to a reduced reaction set of 1597 metabolites and 2902 reactions. We took this reduced set of $N=2902$ reactions as the "global reaction set" or "reaction universe" for our study.

The *E. coli* metabolic model iJR904 contains 931 reactions which occur in the hybrid database of 5870 reactions from which we derived our global reaction set [26]. After having excluded the 2968 blocked reactions, the global reaction set still contains 831 reactions specific to *E. coli*. We consider this space of reactions to be the *E. coli* metabolic genotype.

**Phenotypes and viability**

In general, *in silico* metabolic studies take a metabolic network's "fitness" in a given chemical environment to be proportional to the maximum biomass growth flux the network can attain. The metabolic phenotypes we consider here are conceptually simpler: they regard only viability. Specifically, we consider a genotype to be "viable" in a given chemical environment if and only if its maximum biomass flux is nonzero. That is, the genotype is viable if it can synthesize all biomass compounds, regardless of the synthesis rate. Otherwise, we consider the genotype to be non-viable. We use FBA and the *E. coli* biomass composition mentioned above to determine viability. If a genotype with $n$ reactions in the space $\Omega(n)$ is viable, we say it belongs to the viable space $V(n)$. Thus, the viable space $V(n)$ is a subspace of $\Omega(n)$.

**Chemical environments**

This work is concerned with viability of metabolic network genotypes in a given, well-defined chemical environment or medium. Specifically, we consider only minimal environments that contain a limited amount of a carbon source, along with unlimited amounts of the following inorganic metabolites: oxygen, water, protons, sulfate, ammonia, pyrophosphate, iron, potassium and sodium. The work presented here focuses on the three carbon sources: glucose, acetate and succinate, but we have also investigated properties using four other carbon sources.

**Essential and super-essential reactions**

We call a reaction essential for a given viable genotype in a given chemical environment if its elimination ("knock-out") renders the genotype non-viable in that environment. We call a reaction "super-essential" for this environment if it is essential for the genotype containing all reactions in the global reaction set. We note that a reaction that is super-essential in any one environment must be essential for *every* viable genotype. We have determined the set of super-essential reactions for the different minimal environments we study here. In particular, the glucose and succinate environments have the same 99 super-essential reactions, while acetate's super-essential reactions are those same 99 plus one more. As we describe below, knowledge about these super-essential reactions allows us to increase the efficiency of our sampling of genotype space to estimate the size of the viable space $V(n)$ at much smaller $n$ than would be possible otherwise.

**Estimating the fraction of viable genotypes**

For most of our analysis, we are interested in the properties of genotypes belonging to the viable space $V(n)$. One of the most basic questions is whether the constraint of being viable is "severe", *i.e.,* whether a random genotype in $\Omega(n)$ has a tiny probability of being viable. In principle, one could sample genotypes (bit strings) in $\Omega(n)$ at random and determine for each such genotype whether it is a member of the viable space $V(n)$. In a large enough sample of such genotypes, the fraction of viable genotypes provides an estimate of the fraction of viable genotypes. Unfortunately, for our genotype space, these fractions are extremely small and thus not measurable unless the number $n$ of reactions in a genotype is above 2700. To avoid this problem, we have implemented an algorithmic trick where the probability of being viable is represented as a product of two factors that can be computed separately. The first factor is the probability that a genotype of a given size contains all super-essential reactions. This factor is of a simple combinatorial nature (it is a ratio of two binomial coefficients) and thus can be computed analytically. Specifically, for a given number of $n$ reactions and $s$ super-essential reactions, this factor is given by $B(N-s,n-s)/B(N,n)$ where $B(p,q)$ is the binomial coefficient $p! / [ q! (p-q)! ]$. The second factor is the probability that a genotype is viable given that it contains all super-essential reactions. To estimate this second factor, we use the following procedure to generate $10^6$ random genotypes for a given value of $n$. We first include all reactions in the super-essential set and then randomly include non-super essential reactions until the genotype contains exactly $n$ reactions. The fraction of these $10^6$ genotypes that are viable gives us the estimate of the second factor. This procedure reduces the computational complexity of our sampling problem and allows us to provide accurate estimates of the size of $V(n)$ for a broad range of reaction numbers $n$.

**Aspects of the MCMC algorithm and associated analyses**

In our direct sampling of $V(n)$ at $n>2000$, we exploited the presence of super-essential reactions. This trick can also be used for the MCMC sampling. Specifically, since all genotypes in $V(n)$ contain all super-essential reactions, we can force these reactions to always be present. As a result we can impose that a reaction swap does not involve any super-essential reaction. This procedure enhances the efficiency of the MCMC sampling. Most importantly, it increases the acceptance rate.

In general, the genotypes produced by the MCMC method will have some memory of the starting genotype, although this memory fades with the number of steps performed. (Such memory erasure must occur if the asymptotic distribution is uniform in the accessible space.) In the same vein, the successive genotypes in our Markov chains are strongly correlated, since they differ at best by an exchange of one reaction pair. These correlations typically decrease exponentially with the number of steps, and the associated time scale $\tau$ -- the auto-correlation time of the chain -- can be estimated empirically [37]. We have done so for three different environments, as illustrated for one value of $n$ in Fig. S3. The figure shows that the distance to the initial genotype grows and then quickly saturates, with a characteristic time scale $\tau$ of about two thousand swaps (for the value of $n$ illustrated in the figure), regardless of the environment. Note that $\tau$ gives both the characteristic time to "forget" the starting genotype and the time to go from one genotype to a nearly independent one. This estimation of $\tau$ motivates the following procedure, which we followed for all reaction numbers and environments. Beginning with the initial genotype, we first carried out $10^5$ Markov Chain steps to erase the memory of the starting genotype. After this initial phase, we continued the MCMC procedure to sample the genotype network. During this phase, it is not useful to keep all of the genotypes produced because they are strongly correlated. We thus saved only every $1000^{th}$ genotype generated, and ran the Markov chain for a total of $10^6$ steps, leading to one thousand saved genotypes. These 1000 genotypes are a random sample of viable genotypes in $V(n)$ for a given chemical environment. We repeated the procedure for multiple values of $n$ ($n$=300 to 2800) and for all three minimal environments. We also implemented this procedure for obtaining samples of genotypes that were viable on multiple environments.

The sample produced by the MCMC is unbiased if the memory of the initial genotype is absent (which is why we use a large number of transitions before saving genotypes). But the successive genotypes are to some extent correlated. This does not affect the measurement of averages. However, when estimating statistical errors associated with sampling, it is necessary to take into account that the different genotypes are not independent. The standard procedure to address this problem involves jackknife calculations of the error bars [38]. Briefly, in this approach the ordered data (consisting of the sequence of saved genotypes) is analyzed, the mean of the observable of interest is extracted, and then the jackknife estimate of the statistical error is obtained using variable window sizes. The window size determines how successive data values are regrouped [38]; it must be much smaller than the total sample size, and one can check that the resulting error estimates are insensitive to the choice of this window size. For example, we have used this procedure to estimate errors in the distribution of mutational robustness. For each robustness bin, we determined the number of genotypes of the sample that fall in that bin; and the jackknife estimate gave the error bar on the height of the bin. In practice, mutational robustness has negligible correlations on the time scale of our successively saved genotypes, and thus the jackknife estimates are very similar to those obtained when ignoring correlations.

A rule of thumb for MCMC sampling is that the algorithm is efficient if the autocorrelation time of any observable is not too long. If our sampling disregarded viability, that is if one were sampling $\Omega(n)$ rather than $V(n)$, then one would need of the order of $n$ reaction swaps to generate an independent genotype, leading to $\tau$=$n$. Clearly, when sampling $V(n)$, a proposed swap is refused whenever the modified genotype is not viable. If $A$ is the acceptance rate of swaps, then a simple expectation based on this observation is $\tau = n/A$. In the regime we studied ($n>300$), $A$ is greater than 0.2, as illustrated in Fig. S4 in Additional File 1. In consequence, $\tau$ should be of the order of a few thousand steps. This is what we found (*cf.* Figure S3 in Additional File 1). In sum, the MCMC method is less efficient in $V(n)$ than in $\Omega(n)$, but it is more than adequate for our purposes. Only for genotypes with $n<300$ would the MCMC approach become unacceptably slow, but in this case a much more fundamental problem would be that there may be no genotypes at all in $V(n)$.

### Genotype-specific "essential" and "blocked" reactions

For a randomly sampled genotype viable in a given minimal environment, we can determine both the subset of reactions present that are essential for that genotype in this environment and the subset of reactions that are blocked in this environment. A reaction is blocked for a given genotype and environment if no strictly positive flux through it can occur under steady state conditions [31-32]. In other words, this genotype can have non-zero biomass growth flux only if the flux through this reaction is zero. Such a reaction can be considered as "afunctional" for this genotype and environment.

### Clustering analysis

Our genotypes are bit strings; as such, they can be thought of as vectors in an $N$ dimensional space. Given a list of such vectors, one can use Principal Component Analysis (PCA) to analyze clustering of genotypes. We have also implemented a hierarchical clustering of such strings; to do so, we first construct the list of distances between each genotype, using the Hamming distance. Then we can hierarchically assemble the two nearest clusters into a new (larger) cluster, and repeat this procedure recursively. At each step, it is necessary to have a definition of the distance between two clusters: we take this distance to be the mean distance of members of different clusters. The PCA analysis has been performed using the function *princomp* in MATLAB 7.7. The hierarchical clustering analysis has been performed using functions *pdist*, *linkage* and *dendrogram* in MATLAB 7.7.

## Acknowledgements

We thank Monique Bolotin-Fukuhara for discussions. AS thanks the Federation of European Biochemical Societies (FEBS) for a short term fellowship to visit OCM. AW and JR acknowledge support from the Swiss National Science Foundation as well as from the YeastX project of SystemsX.ch, and JJ acknowledges support from the Volkswagen Foundation. OCM acknowledges support from the Agence Nationale de la Recherche, Metacoli grant ANR-08-SYSC-011. We would also thank the anonymous reviewers for their constructive comments on the manuscript.

# Figures

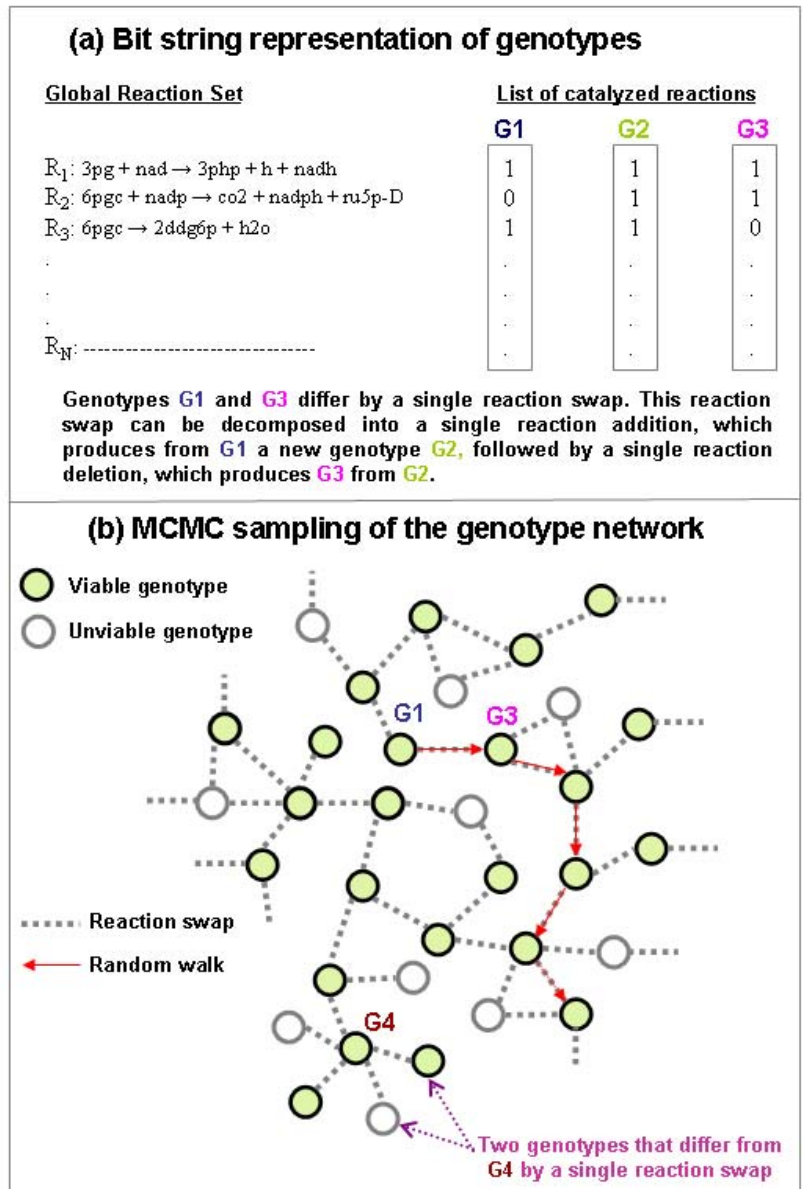

**Figure 1: Genotypes and the sampling of genotype networks. a)** Genotypes are subsets of reactions in a global reaction set. Since the global set has $N$ reactions, a genotype having $n$ reactions can be represented by a bit string of length $N$ with $n$ entries equal to 1 and all others equal to 0. **b)** At given $n$, the set of *viable* genotypes forms a "genotype network" that is a tiny fraction of the whole space. MCMC allows one to sample that tiny fraction by generating a random walk among viable genotypes, going from one viable to another by performing a reaction swap. If a swap leads to a non-viable genotype, the walk stays at the previous genotype for that step and then the process is repeated. The advantage of using reaction swaps in our approach is that it leaves the number of reactions constant over time.

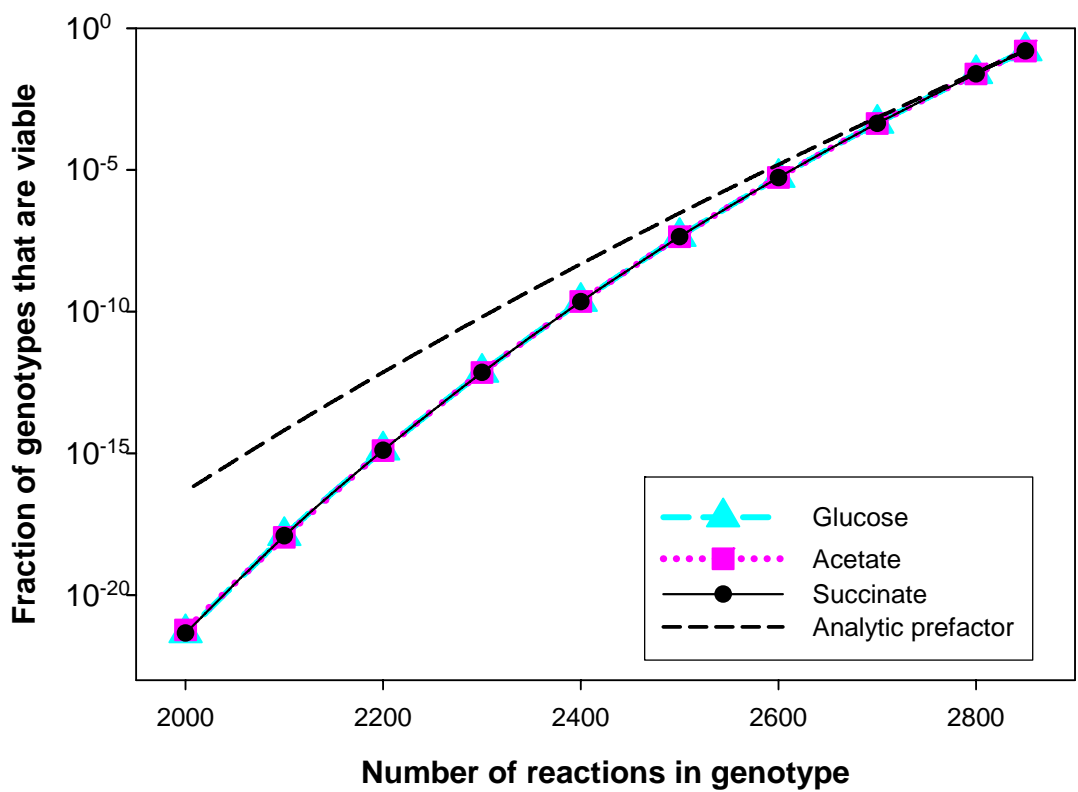

**Figure 2: The space of viable genotypes gets rarefied with decreasing *n*.** The horizontal axis shows the number *n* of reactions in a genotype and the vertical axis shows, on a logarithmic scale, the estimated fraction of random genotypes contained in the viable space *V(n)* for three different chemical environments, glucose, acetate and succinate, respectively. The estimated fraction of viable genotypes shown in this figure has been obtained as a product of two terms. The first term is an analytic prefactor which gives the fraction of genotypes in *Ω(n)* containing all super-essential reactions. The black dashed curve shows this analytic prefactor as function of *n*. The second term, that we estimated numerically, is the probability for a genotype to be viable given that it has all super-essential reactions. The numerical estimation of the fraction of viable genotypes in *Ω(n)* becomes extremely difficult for *n*<2100 reactions even with the decomposition method.

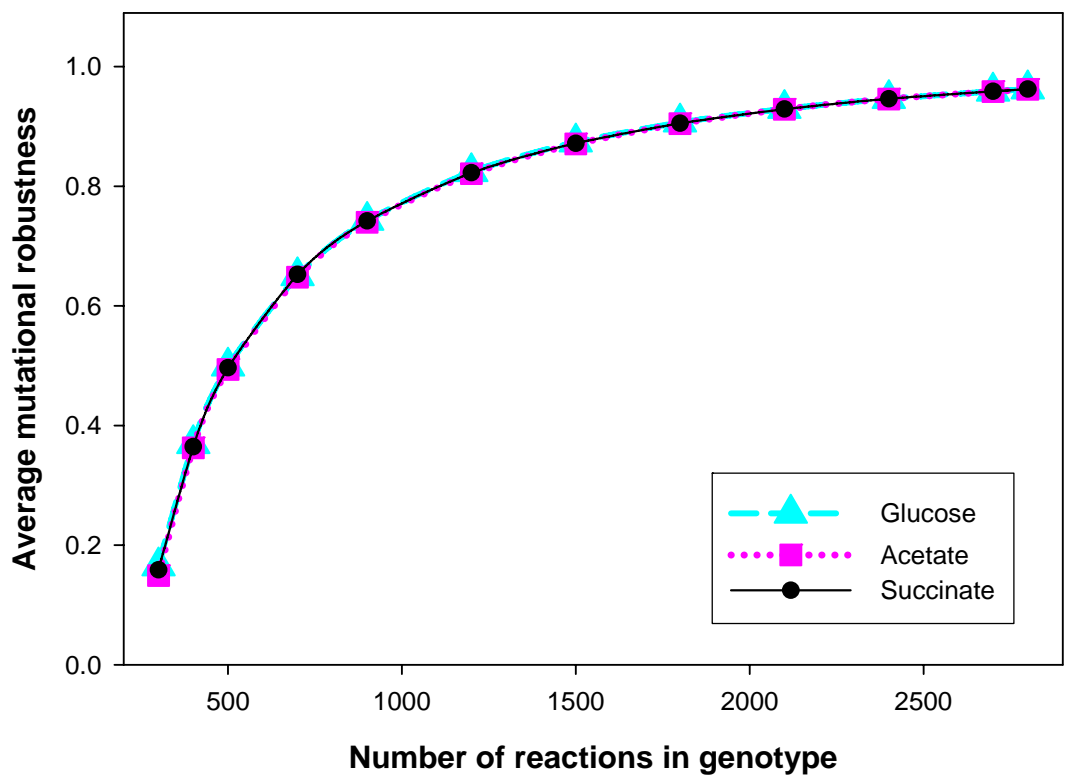

**Figure 3: Mutational robustness $R_\mu$ increases with *n*.** The horizontal axis shows the number *n* of reactions in a genotype and the vertical axis shows the average mutational robustness of sampled genotypes in *V(n)* for three environments (glucose, acetate and succinate) as a function of *n*. Mutational robustness $R_\mu$ is defined as the fraction of non-essential reactions in a viable genotype. It can be seen that $R_\mu$ is comparable to the acceptance rate *A* (*cf.* Figure S4 in Additional File 1). Even for *n*=300, $R_\mu$>0.15, indicating that robustness is not too low for an efficient MCMC sampling.

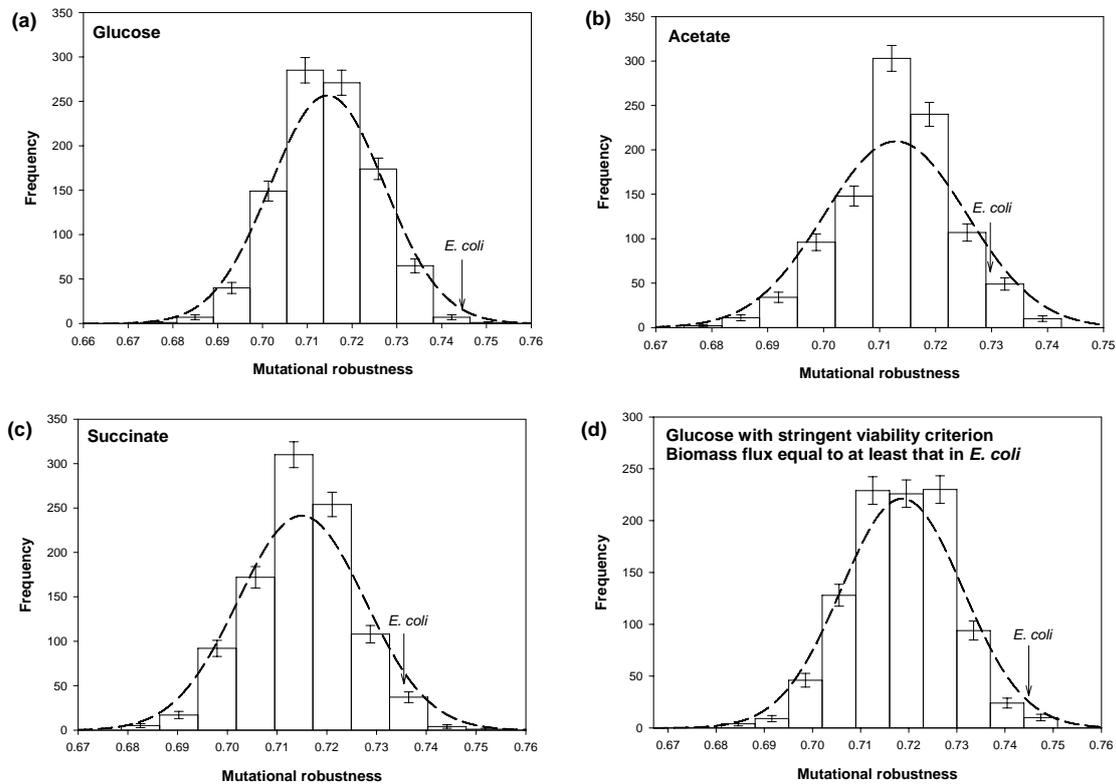

**Figure 4: Narrow distribution of mutational robustness $R_\mu$ for genotypes in $V(n)$.** The horizontal axis shows the mutational robustness $R_\mu$ and the vertical axis shows the frequency of genotypes with the corresponding value of $R_\mu$ in a random sample of 1000 genotypes viable in **a)** glucose, **b)** acetate and **c)** succinate environment with the viability constraint taken as having strictly positive biomass flux. **d)** The distribution of mutational robustness $R_\mu$ in 1000 random viable genotypes for the glucose environment with viability constraint taken as biomass flux at least as large as the *in silico E. coli* biomass flux [24]. In all subfigures, we have shown the normal distribution with sampled mean and theoretically predicted variance as a dashed black curve (see text for details). The figures show that there is very little variation in $R_\mu$ across random viable genotypes and the normal distribution agrees well with the sampled distribution. In all cases, the number $n$ of reactions is equal to that in *E. coli* ($n$=831).

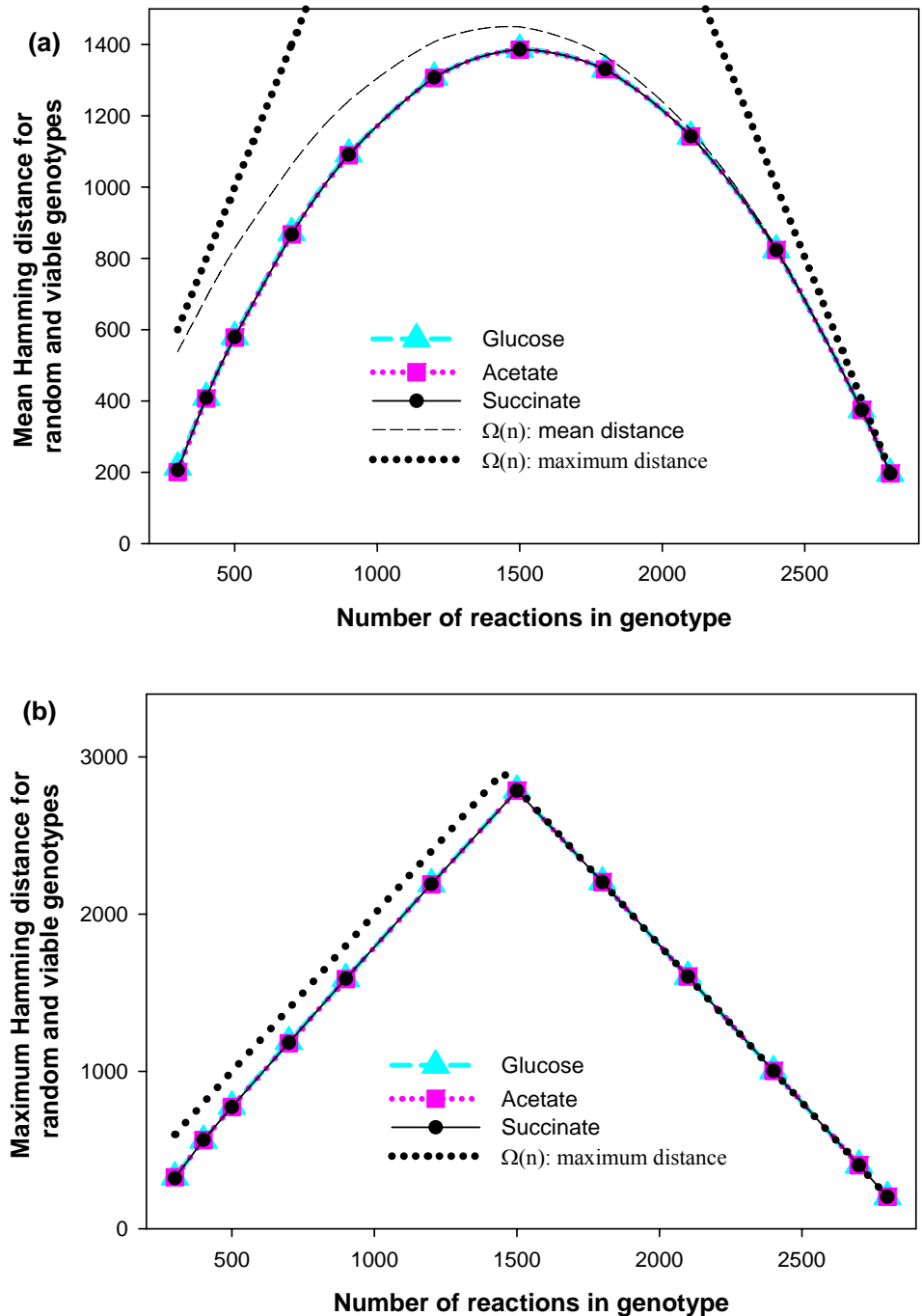

**Figure 5: a) Mean Hamming distance between viable genotypes.** The horizontal axis shows the number *n* of reactions and the vertical axis shows the mean Hamming distance between random genotypes in the set *V(n)*, and for a comparison, in *Ω(n)*. Data is shown for three different environments (glucose, acetate and succinate). For each *V(n)*, we computed the data using all pairwise distances between our 1000 sampled genotypes. The dashed black curve shows mean Hamming distance for the case of *Ω(n),* which can be computed analytically and is given by *2 n(N-n)/N.* The viability constraint affects the mean Hamming distance significantly only at low values of *n*. We also show the maximum Hamming distance in *Ω(n)* (dotted black line). **b) Maximum Hamming distance between viable genotypes.** The horizontal axis shows the number *n* of reactions and the vertical axis shows the maximum Hamming distance between genotypes in the set *V(n)* and, for comparison, in *Ω(n)*. Data is shown for three different environments. For each *V(n)*, we computed the data through a modified two-headed MCMC approach (see text for details). The dotted black curve shows the maximum distance in *Ω(n)* which can be computed analytically and is given by *2n* for *n<N/2* and *2(N-n)* for *n>N/2*. Again, the viability constraint affects the maximum Hamming distance significantly only at low values of *n*.

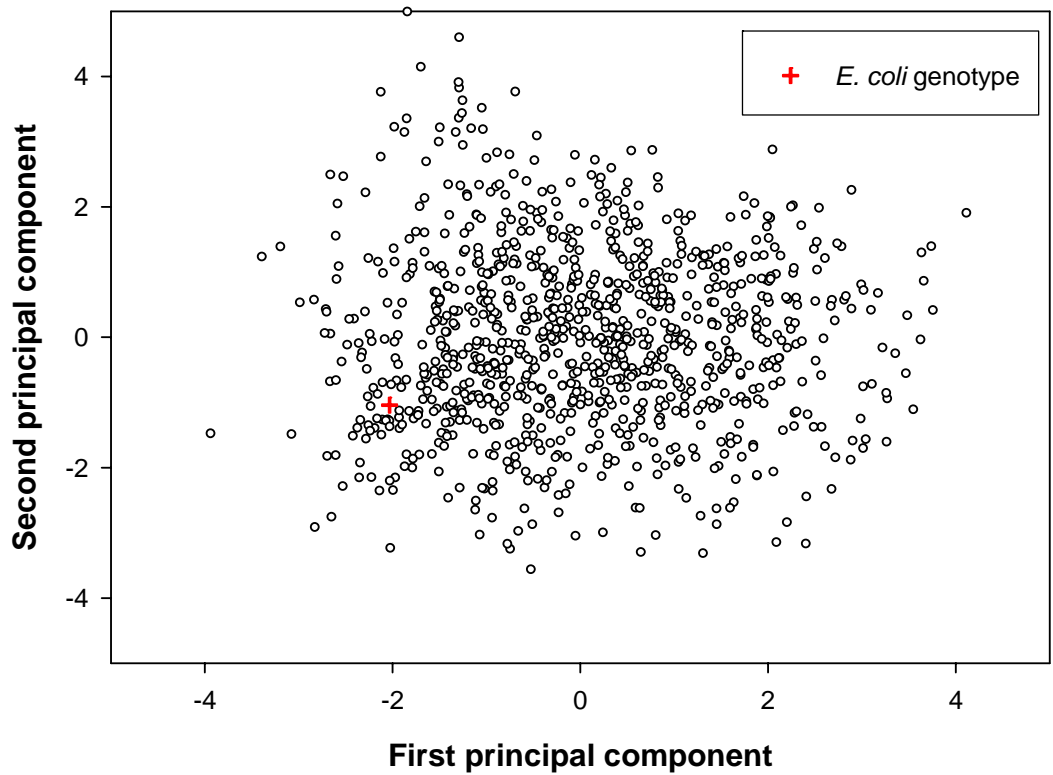

**Figure 6: Principal component analysis of sampled genotypes in *V(n).*** The figure shows the first two principal components for the randomly sampled genotypes that are viable in the glucose chemical environment with *n* equal to that of *E. coli*. The horizontal axis shows the first principal component, and the vertical axis shows the second principal component. The figure additionally shows in red the *E. coli* genotype in this space. The first two components account for less than 0.8% of the variance in the data. The analysis suggests that there are no multiple clusters.

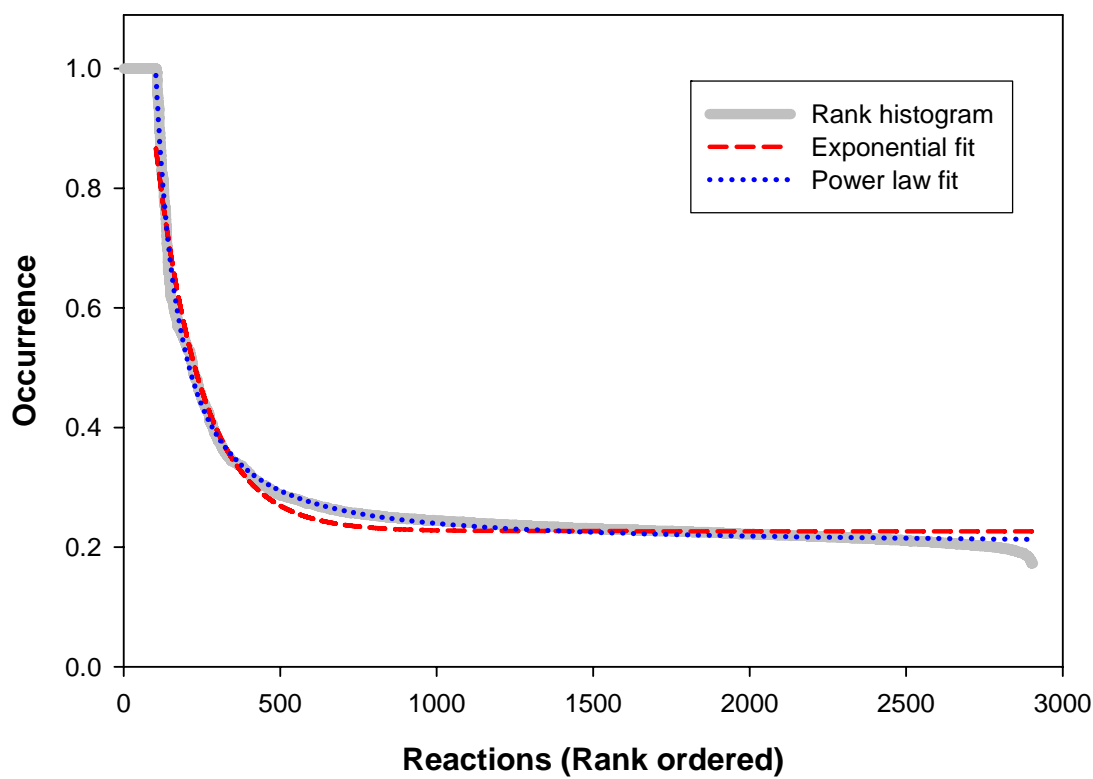

**Figure 7: Rank histogram of reactions based on their occurrence.** The horizontal axis shows the 2902 reactions in our global reaction set ordered based on their rank. The vertical axis shows the occurrence of each reaction in 1000 randomly sampled genotypes that are viable in a glucose environment, with $n$=831 equal to that of *E. coli*. The occurrence of a reaction is given by the number of sampled genotypes containing the reaction, divided by the sample size of 1000. Reactions have been ordered based on their decreasing occurrence, with rank "1" corresponding to reaction with highest occurrence. All super-essential reactions have an occurrence of 1.0 and contribute to the horizontal plateau on the left of the rank histogram. We also see a larger plateau on the right, with an associated occurrence value of approximately 0.2. The two plateaus are connected by a degraded slope, corresponding to reactions that see their occurrence decrease continuously from 1.0 to 0.2. To study this region of the rank histogram, we have fitted it to two classes of functions: a constant plus an exponential (*f(r)=a+bexp[-cr]*; *a*=0.2267, *b*=1.293, *c*=0.006827) and a constant plus a power function (*f(r)=a+b/r^c*; *a*=0.2052, *b*=458.9, *c*=1.375). Based on the coefficient of determination $R^2$, the power law provides a better fit ($R^2$=0.9645 for the exponential and $R^2$=0.9927 for the power).

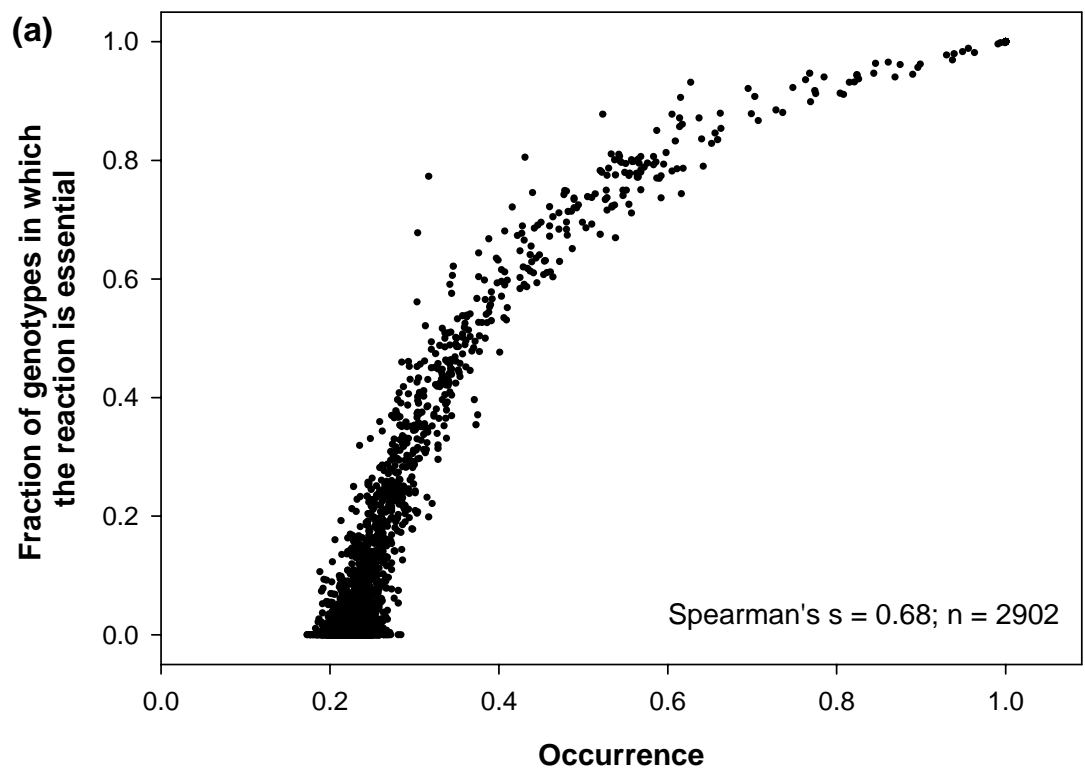
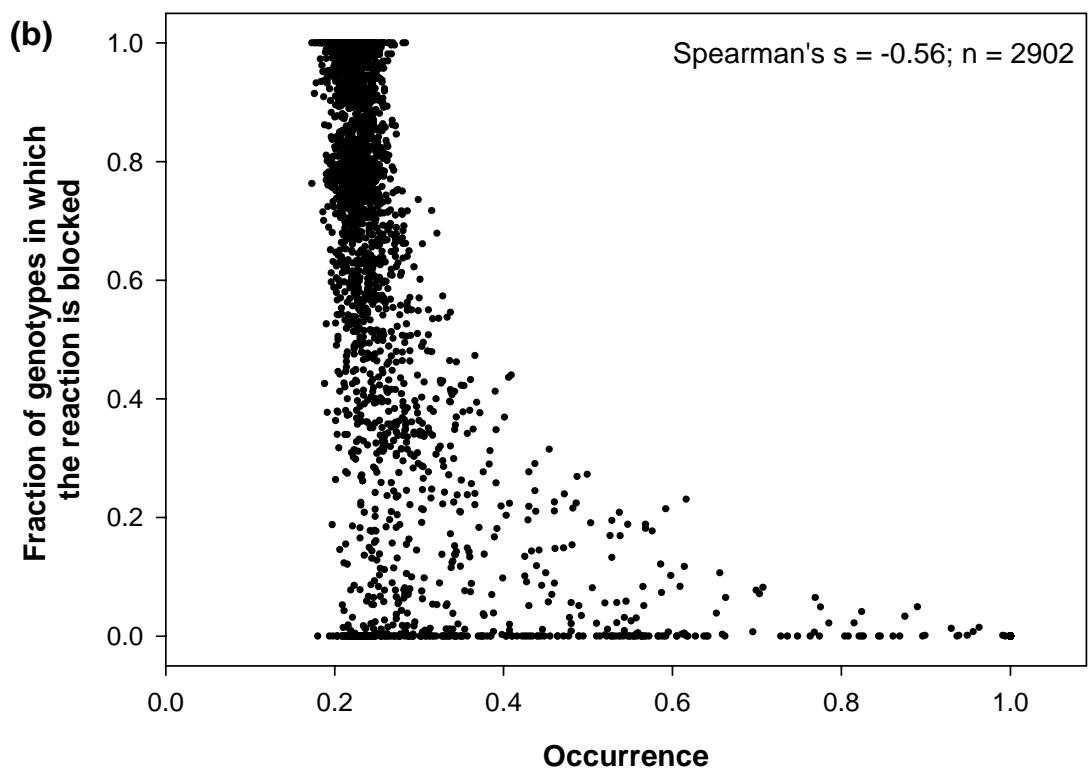

**Figure 8: Frequency of reaction use is correlated with essentiality. a)** The horizontal axis shows the occurrence of a reaction and the vertical axis shows the frequency with which the reaction is essential in random viable genotypes. A strongly positive correlation exists between the occurrence of a reaction and the frequency with which it is essential. **b)** The horizontal axis shows the occurrence of a reaction, and the vertical axis shows the frequency with which the reaction is blocked. Data is based on 1000 randomly sampled genotypes viable on glucose minimal environment with $n$=831 reactions, the number of reactions in *E. coli,*.

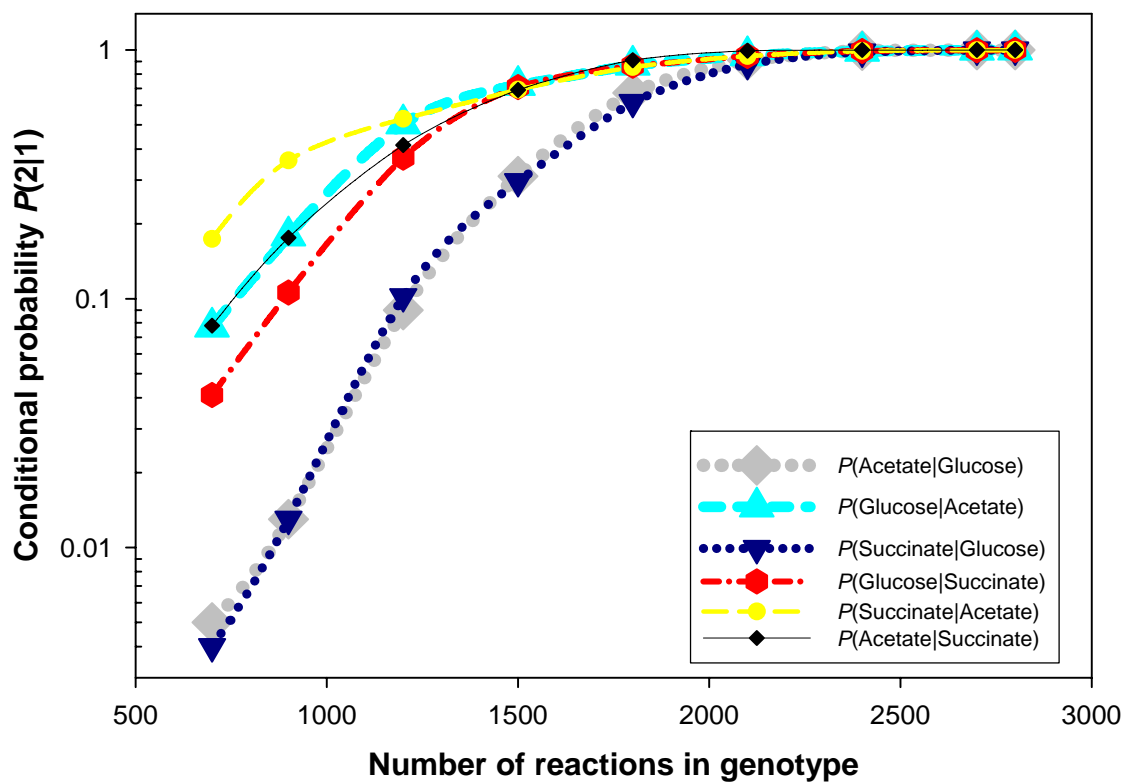

**Figure 9: Conditional probability of a genotype viable in environment 1 to be also viable in environment 2.** The horizontal axis shows the number of reactions in a genotype, and the vertical axis shows the conditional probability *P(2|1)* that a genotype is viable in environment 2 given that it is viable in environment 1. The figure shows this conditional probability for all six ordered pairs for the three different environments, glucose, acetate and succinate.

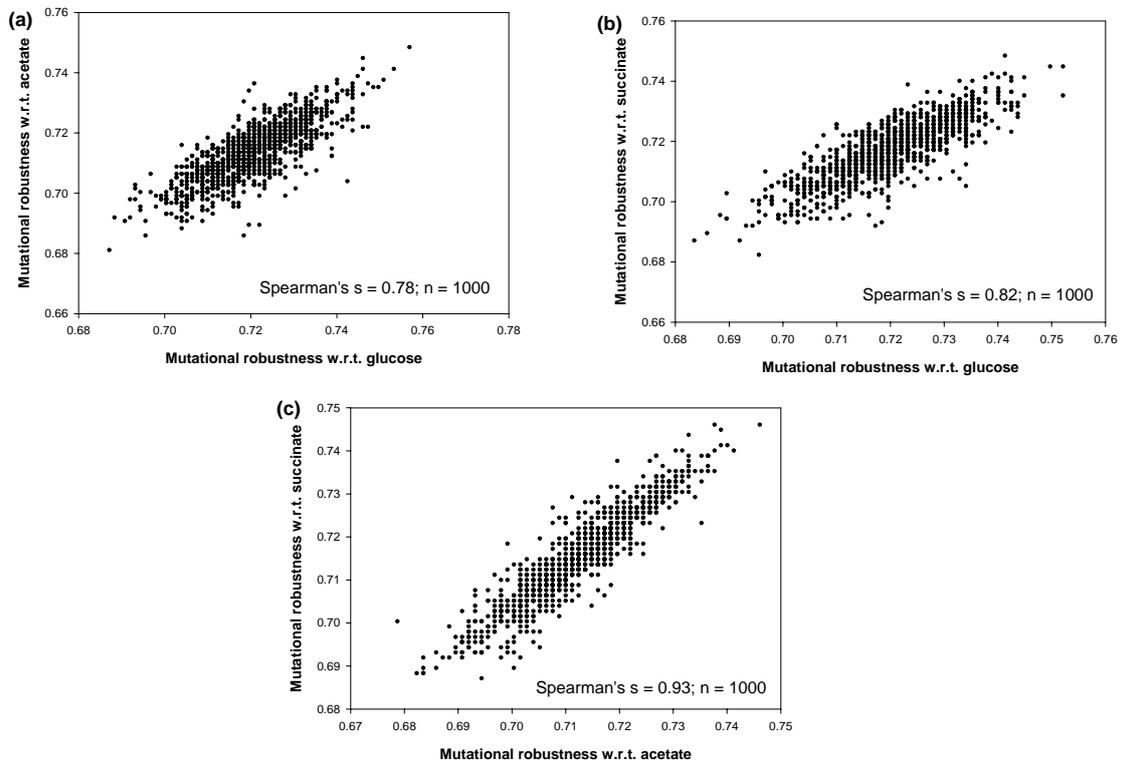

**Figure 10: High correlation of mutational robustnesses in one environment for genotypes that are viable in two environments.** The horizontal axis shows the mutational robustness with respect to a first environment and the vertical axis shows the mutational robustness with respect to a second environment. Data are based on 1000 sampled genotypes with *n*=831 reactions, equal to the number of reactions in *E. coli*. Genotypes are viable in **a)** glucose and acetate, **b)** glucose and succinate, and **c)** acetate and succinate environments. In all three cases, we can see a high positive association between robustnesses for each environment.

**Additional File 1: Supplementary Figures S1 to S14 of**

**"Genotype networks in metabolic reaction spaces"**

Areejit Samal, João F. Matias Rodrigues, Jürgen Jost, Olivier C. Martin, Andreas Wagner

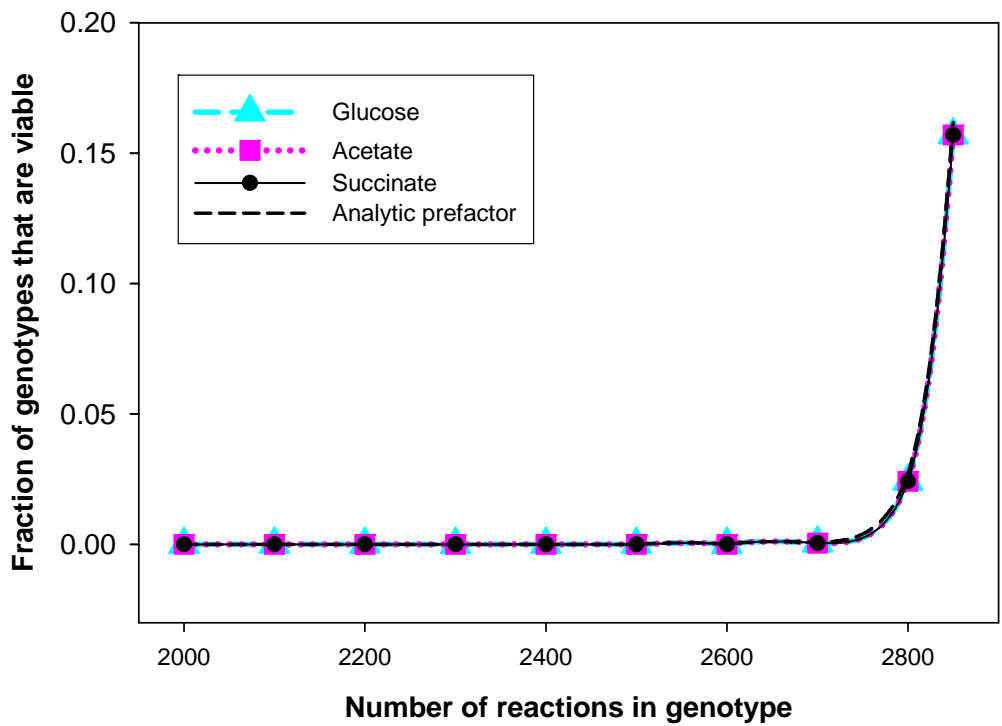

**Figure S1: The space of viable genotypes gets rarefied with decreasing *n*.** The horizontal axis shows the number *n* of reactions in a genotype and the vertical axis shows on a linear scale the estimated fraction of random genotypes contained in the viable space *V(n)* for three different chemical environments, glucose, acetate and succinate, respectively. The black dashed curve shows the analytical prefactor as function of the number *n*.

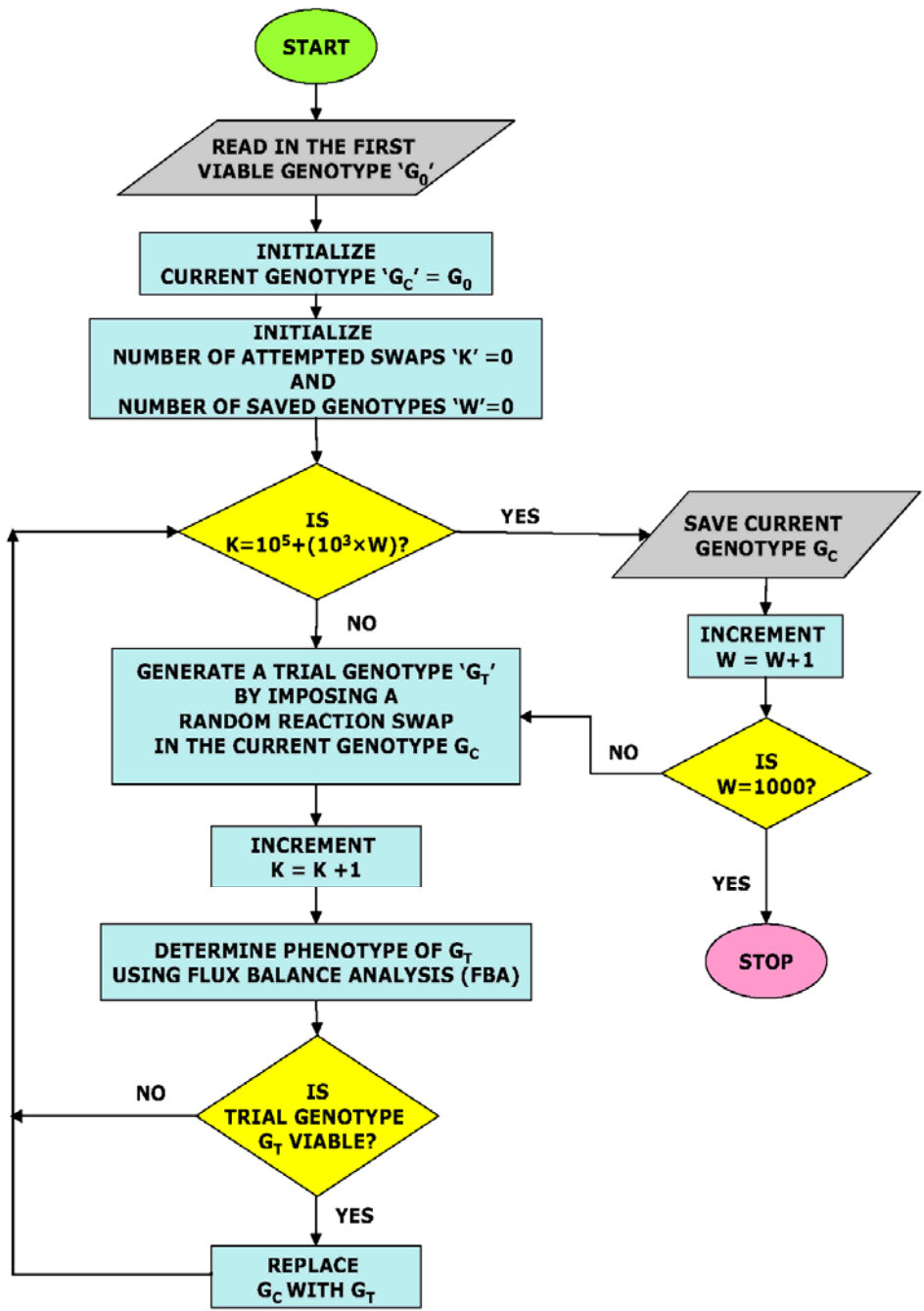

**Figure S2: Flowchart describing the logic of the MCMC algorithm.** To start the Markov chain, a first viable genotype is necessary. Beginning with the initial genotype, we perform $10^5$ Markov Chain steps to erase the memory of the initial genotype. After this initial phase, we continue the MCMC procedure to sample the genotype network and save every $1000^{th}$ genotype generated. We terminate the Markov chain after saving 1000 genotypes. Note that the acceptance rate of swaps should be high enough to obtain a meaningful sample of viable genotypes using this algorithm.

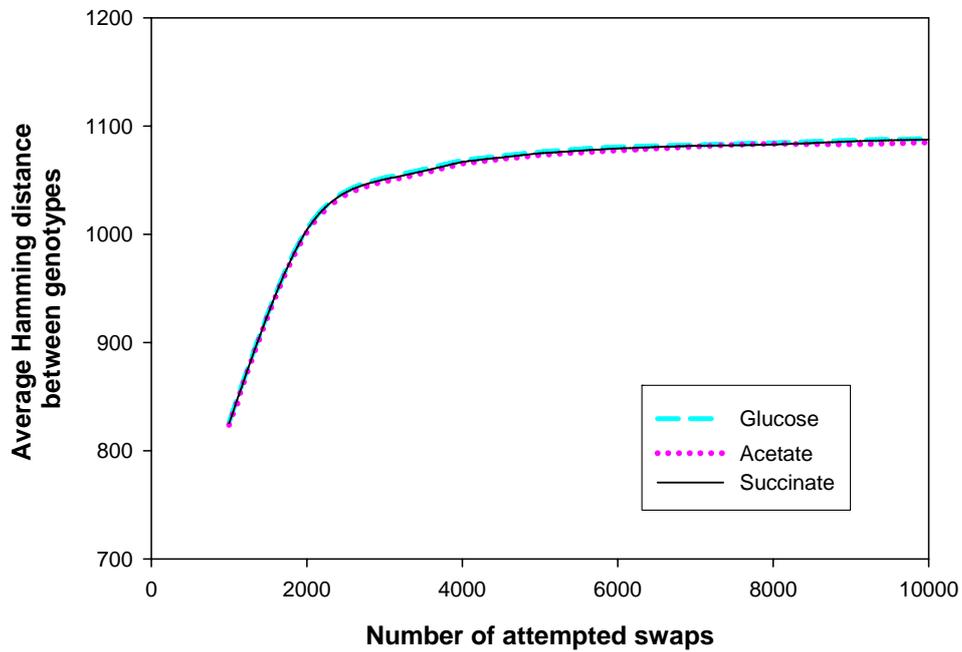

**Figure S3: Autocorrelation time of a typical MCMC run.** The horizontal axis shows the number of attempted swaps in an MCMC run and the vertical axis shows the average Hamming distance between the starting and current genotype for three environments (glucose, acetate and succinate). This distance grows and then saturates with a characteristic time scale $\tau$ of about two thousand attempted swaps for all three environments. Data is shown for $V(n)$ with $n$=900.

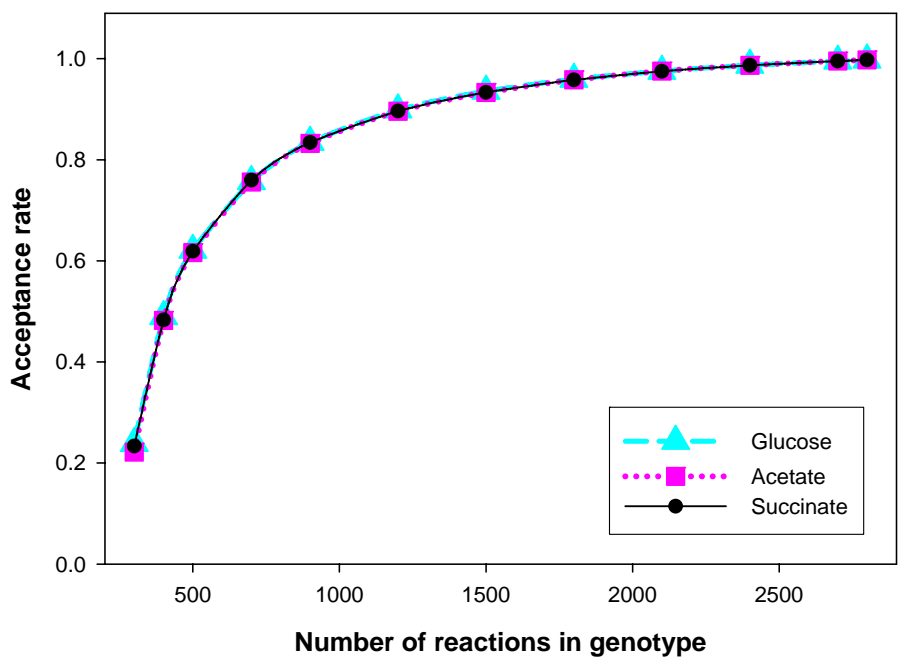

**Figure S4: Acceptance rate of the MCMC procedure.** The horizontal axis shows the number $n$ of reactions in a genotype and the vertical axis shows the acceptance rate $A$ of the MCMC transition steps (reaction swaps). Results are shown for three different environments (glucose, succinate, and acetate). One sees $A>0.2$ for $n=300$ or more which is our regime of interest. Note that during the MCMC sampling of $V(n)$, the super-essential reactions do not participate in the swaps.

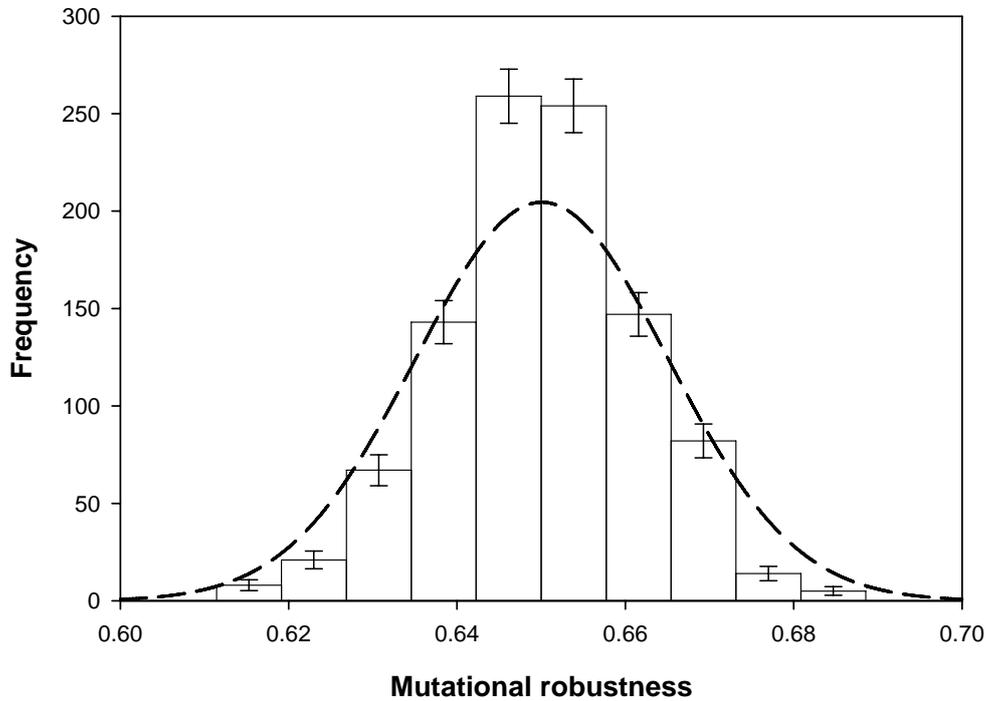

**Figure S5: Narrow distribution of mutational robustness $R_\mu$ for genotypes in $V(n)$.** The horizontal axis shows the mutational robustness $R_\mu$, and the vertical axis shows the frequency of genotypes with the corresponding value of $R_\mu$ in a random sample of 1000 viable genotypes in the glucose minimal environment with the viability constraint taken as having strictly positive biomass flux and the number $n$=700 reactions. We also display the normal distribution with sampled mean and theoretically predicted variance as a dashed black curve (see text for details). The figure shows that there is very little variation in $R_\mu$ across random viable genotypes and the normal distribution agrees relatively well with the sampled distribution.

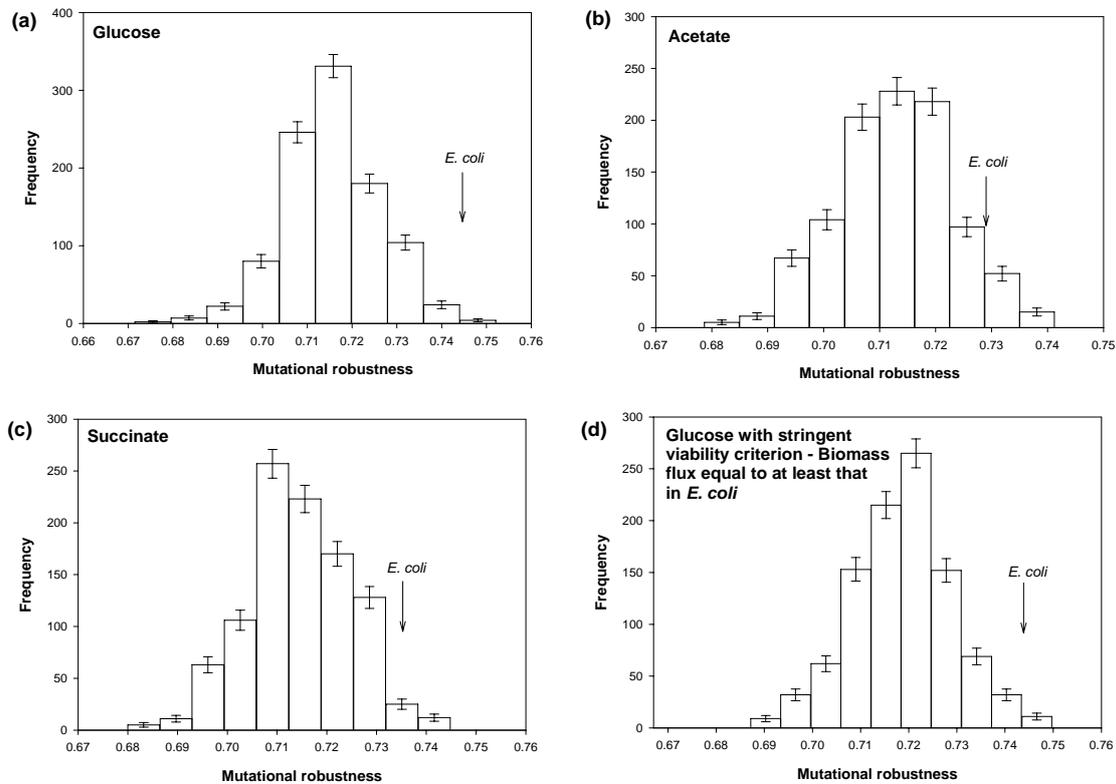

**Figure S6: Narrow distribution of mutational robustness $R_\mu$ for genotypes in $V(n)$ with modified biomass formula.** The modified biomass formula was constructed by starting from the reference one (that of *E. coli*) and randomly perturbing the stoichiometry of each biomass metabolite by up to 20%. The horizontal axis shows the mutational robustness $R_\mu$ and the vertical axis shows the frequency of genotypes with the corresponding value of $R_\mu$ in a random sample of 1000 genotypes viable in **a)** glucose, **b)** acetate and **c)** succinate environment with the viability constraint taken as having strictly positive biomass flux with the modified biomass formula. **d)** The distribution of mutational robustness $R_\mu$ in 1000 random viable genotypes for the glucose environment with viability constraint taken as biomass flux at least as large as the *in silico E. coli* biomass flux with the modified formula. In all cases, the number *n* of reactions is equal to that in *E. coli* (*n*=831). The figure confirms that *E. coli* has atypical mutational robustness even with the modified biomass formula.

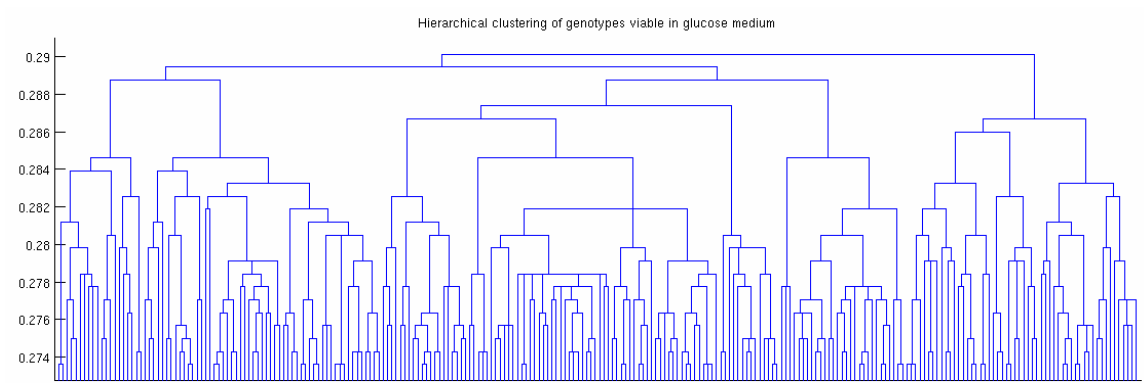

**Figure S7: Hierarchical clustering of genotypes in *V*(*n*).** The figure shows the dendrogram obtained by hierarchically clustering the 1000 sampled viable genotypes for glucose minimal environment with *n*=831 (the value for *E. coli*). Here, Hamming distance was used as the distance measure, followed by average linkage clustering as implemented in MATLAB 7.7. There is no evidence for multiple clusters.

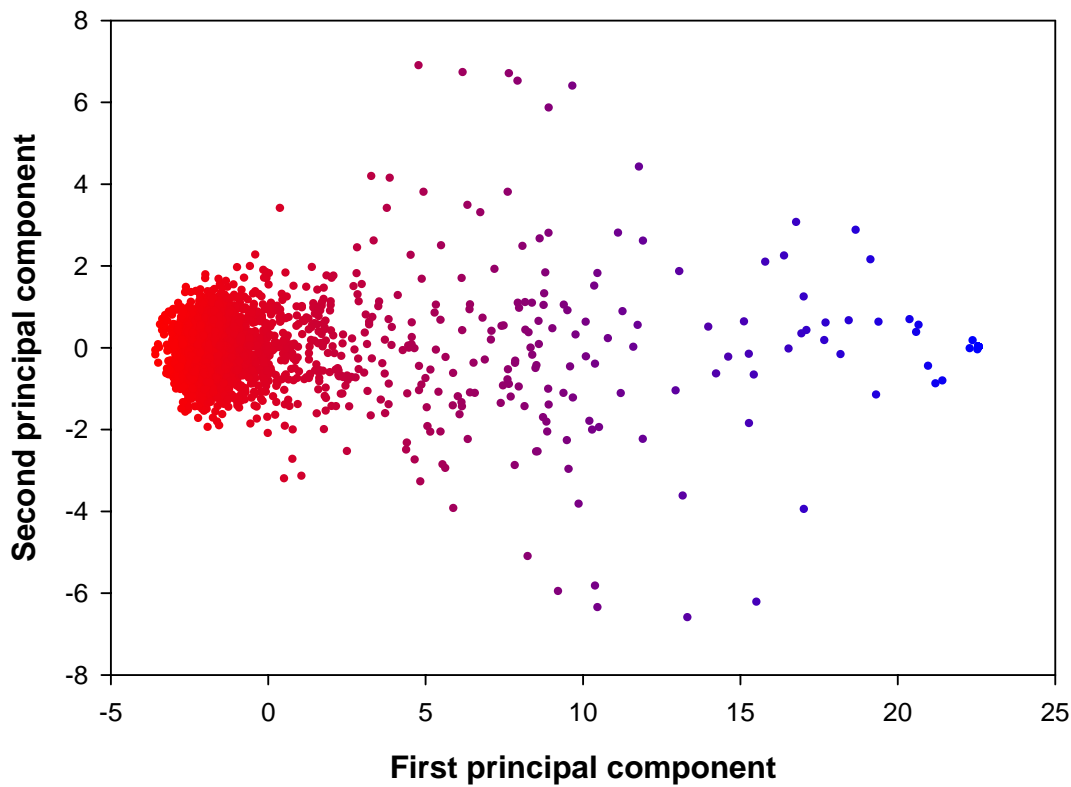

**Figure S8: Principal component analysis of reactions occurring in random viable genotypes.** We have organized the 1000 sampled viable genotypes for glucose minimal environment with *n*=831 (the value for *E. coli*) into a matrix where each row is a bit string associated with one genotype. When we read this matrix one column at a time, we have a bit string of length 1000 that reflects the occurrence of each reaction in the global reaction set in our genotype sample. We subjected these bit strings for reaction occurrence to a principal component analysis. The horizontal axis shows the first principal component and the vertical axis shows the second principal component. We find that the first principal component correlates well with the rank of the reaction. To make the association between the first axis and reaction rank visible, we have colored the reactions according to their rank (red for ranks close to 1, indigo for ranks close to 2902). The data is clearly heterogeneous, resembling a comet with a dense head on the left and a spread-out tail on the right. The comet's head is formed mostly by blocked reactions, while the tail of the comet is enriched in essential reactions.

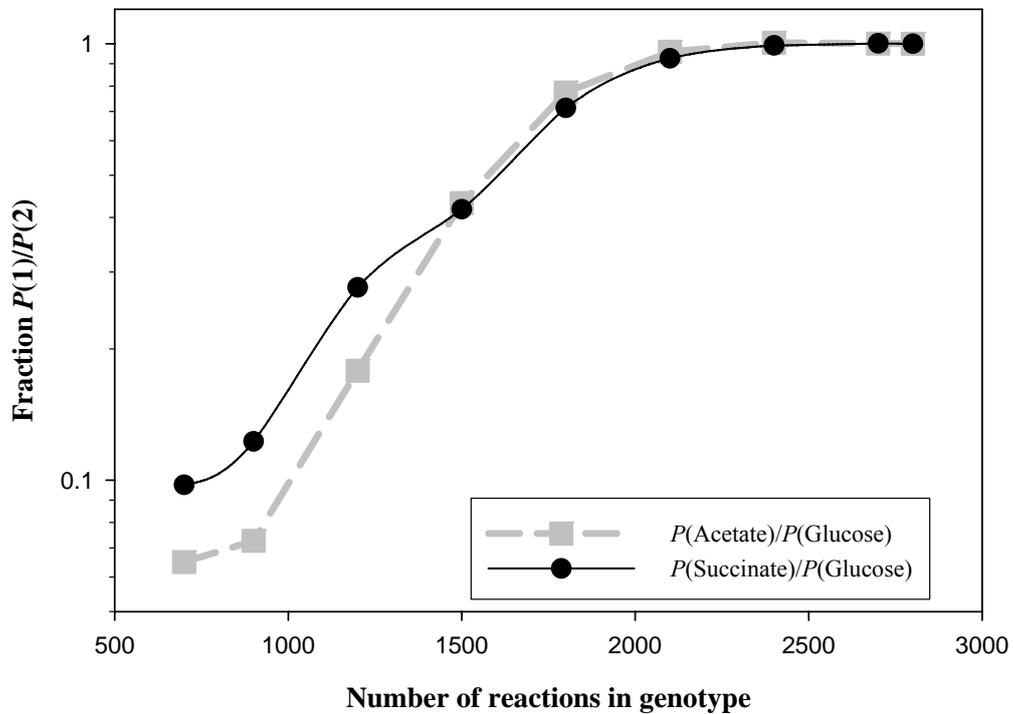

**Figure S9: Relative size of viable spaces for three different environments.** The horizontal axis shows the number of reactions in a genotype, and the vertical axis shows the relative size of the viable space for one environment relative to that for glucose. We here considered the viable spaces for three different environments, glucose, acetate and succinate, and found that the size of the viable space for glucose is greater than that of succinate, which is in turn is greater than that for acetate.

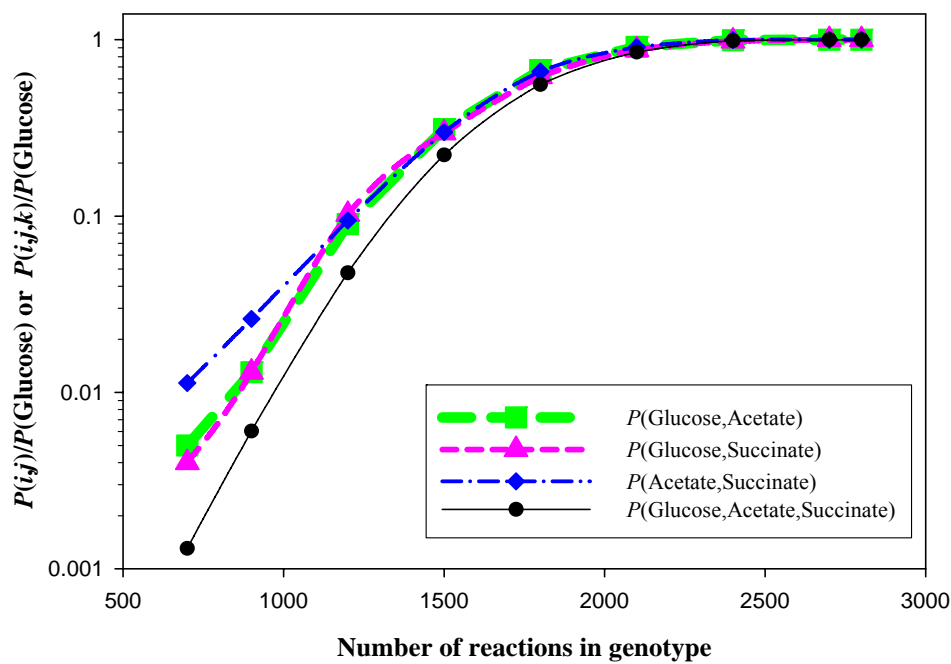

**Figure S10: Probability of a genotype to be viable in multiple environments.** The horizontal axis shows the number of reactions in a genotype and the vertical axis shows the probability that the genotype is simultaneously viable in several environments. We show the data for all three pairs and one triplet of the three environments (glucose, acetate and succinate). The probabilities shown are normalized by the fraction of genotypes that are viable in the glucose environment.

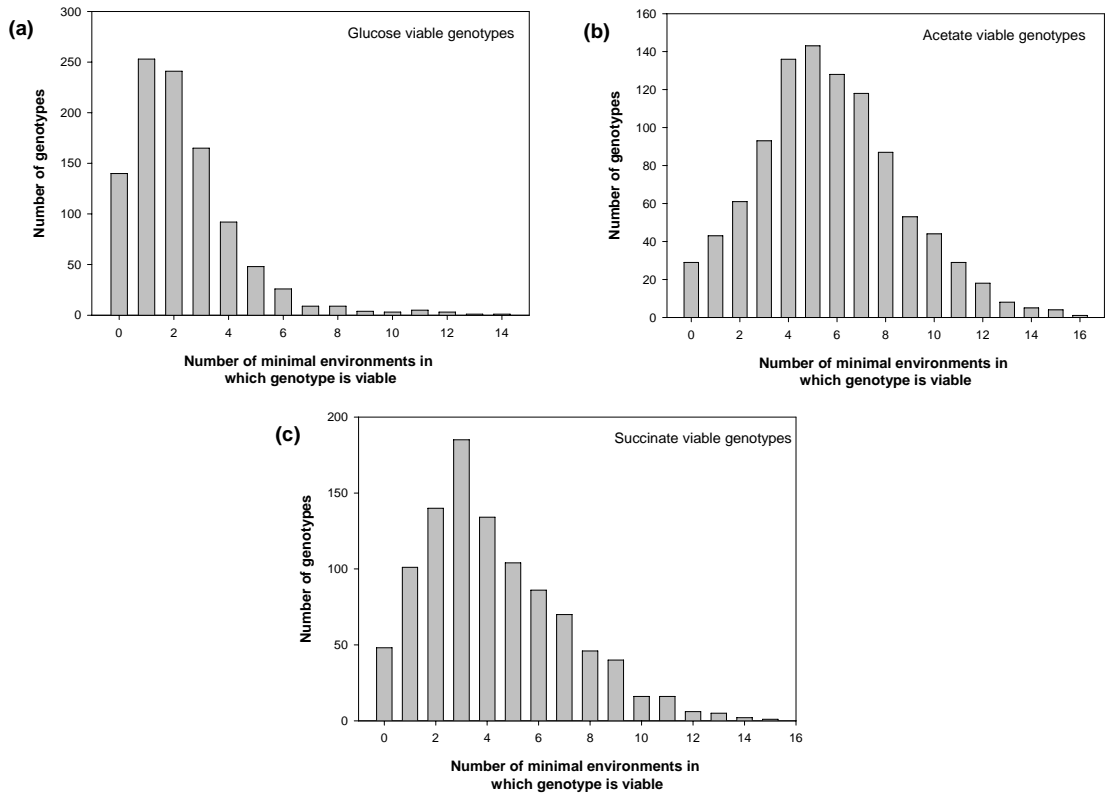

**Figure S11: Histogram of the number of additional environments in which random genotypes are viable.** The horizontal axis shows the number $E$ of environments out of the 88 minimal environments in which a random genotype is viable (apart from the single environment where it was sampled from). The vertical axis shows, as a function of $E$, the corresponding number of genotypes found in our random sample of 1000 genotypes. The 1000 random viable genotypes were sampled to be viable in the **a)** glucose, **b)** acetate and **c)** succinate environment.

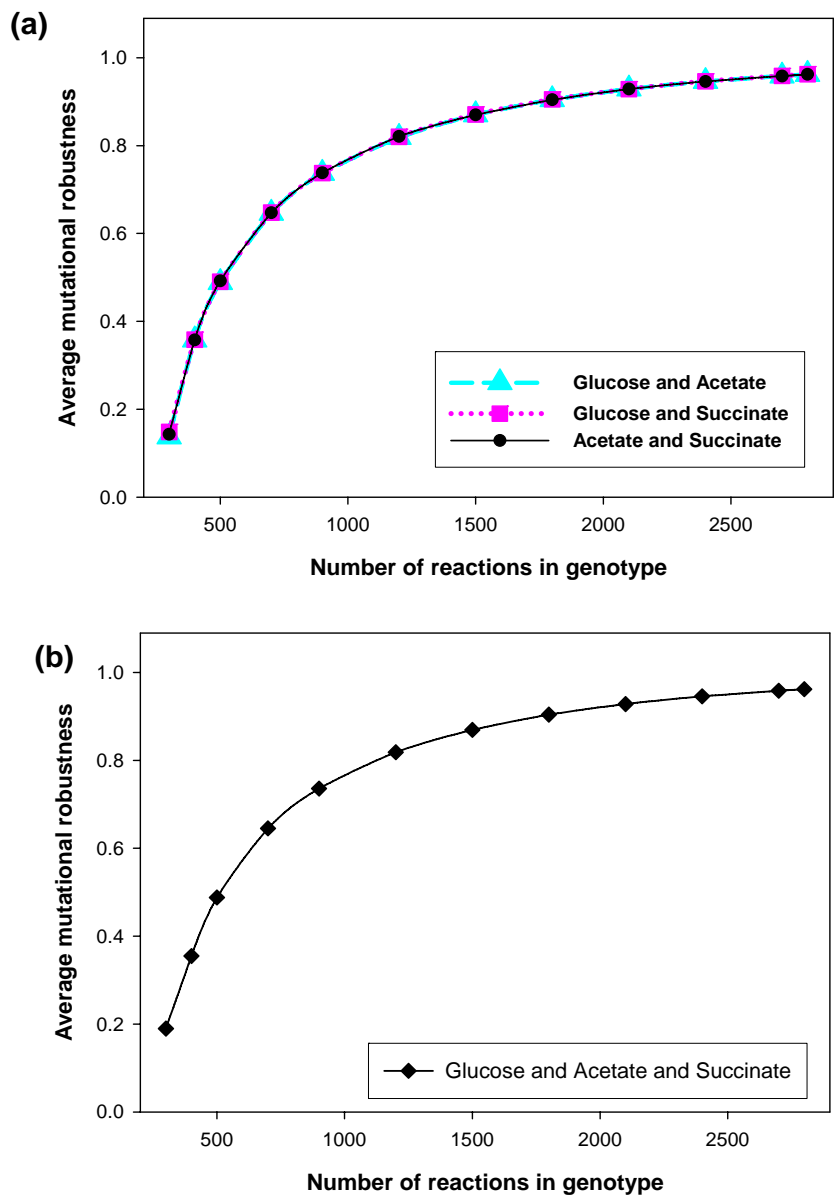

**Figure S12: Mutational robustness $R_\mu$ increases with *n* for genotypes viable in multiple environments. a)** The horizontal axis shows the number *n* of reactions in a genotype, and the vertical axis shows the average mutational robustness of sampled genotypes that are simultaneously viable in two different environments as a function of *n*. The data is shown for sampled genotypes that are viable in all three possible pairwise environment combinations involving glucose, acetate and succinate. **b)** The horizontal axis shows the number *n* of reactions in a genotype and the vertical axis shows the average mutational robustness of sampled genotypes viable in all three environments (glucose, acetate and succinate) as a function of *n*.

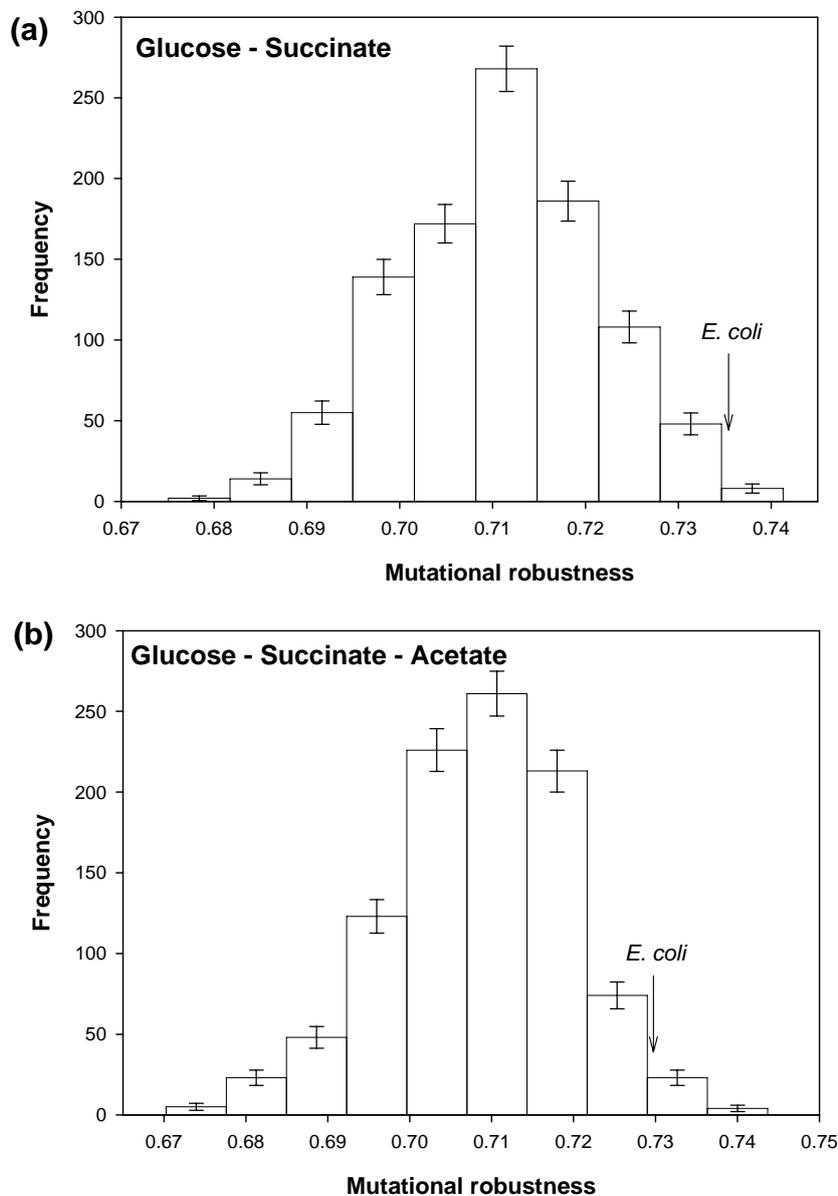

**Figure S13: Narrow distribution of mutational robustness $R_\mu$ for genotypes viable in multiple environments.** The horizontal axis shows the mutational robustness $R_\mu$, and the vertical axis shows the frequency of genotypes with the corresponding value of $R_\mu$ in a random sample of 1000 genotypes that are viable **a)** in the glucose and succinate minimal environments; and **b)** in all three environments (glucose, acetate and succinate). The viability constraint is that of strictly positive biomass flux; the number $n$=831 of reactions is equal to that of *E. coli*. The figures show that there is very little variation in $R_\mu$ across random viable genotypes and that the *E. coli* genotype is an outlier compared to random viable genotypes.

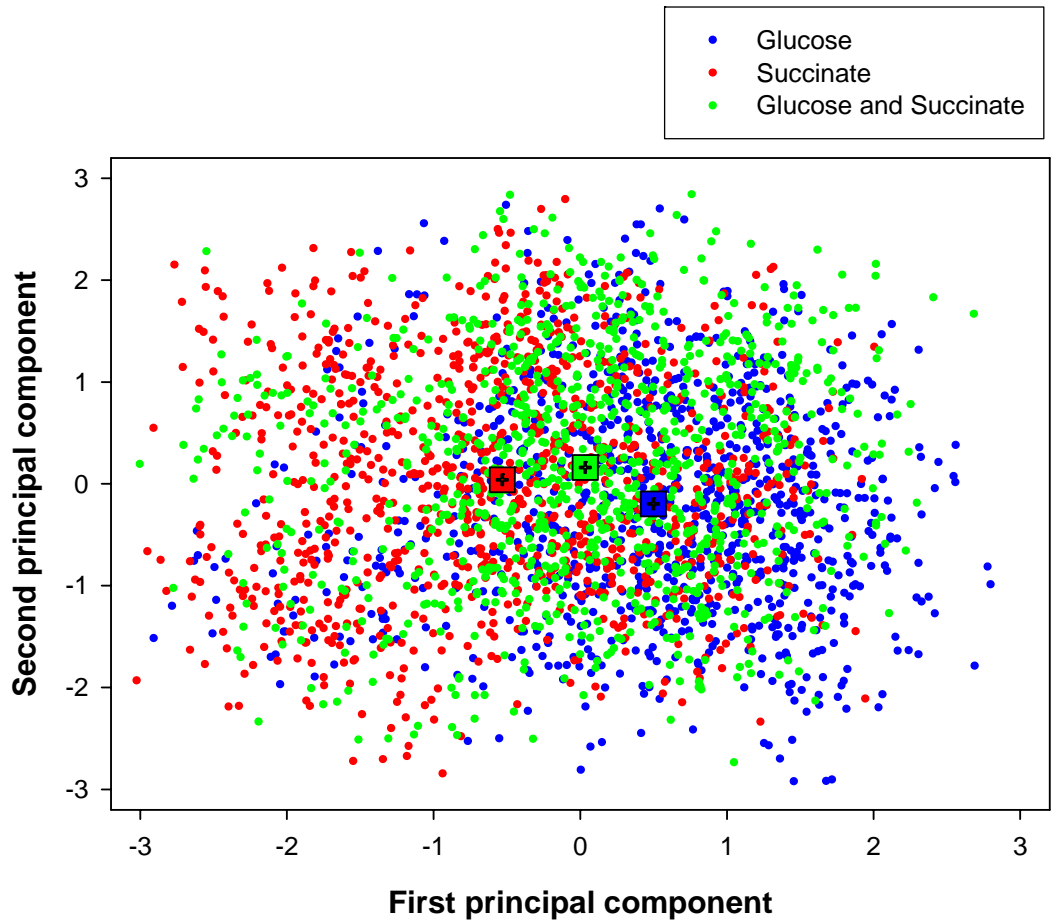

**Figure S14: Clustering of genotypes in *V(n)* for glucose, succinate and both environments**. The figure shows the first two principal components for the randomly sampled viable genotypes in the three considered chemical environments (glucose, succinate and both) with *n* equal to that of *E. coli*. The horizontal axis shows the first principal component, and the vertical axis shows the second principal component. The figure also shows the center of mass of each of the three sets as squares.